\documentclass[reprint,twocolumn,longbibliography, amsmath,amssymb,aps]{revtex4}

\usepackage{graphicx}
\usepackage{pgfplots}
\usepackage{mathtools}
\usepackage{dcolumn}
\usepackage{bm}
\usepackage{float}
\usepackage{xcolor}

\begin{document}

\preprint{APS/123-QED}

\title{Epidemic threshold and localization of the SIS model on directed complex networks}

\author{Vinícius B. M\"uller and Fernando L. Metz}
\affiliation{Physics Institute, Federal University of Rio Grande do Sul, 91501-970 Porto Alegre, Brazil}

\begin{abstract}
  We study the susceptible-infected-susceptible (SIS) model on directed complex networks within the quenched mean-field approximation. Combining results from
  random matrix theory
  with an analytic approach to the distribution of fixed-point infection probabilities, we derive the phase diagram and show
that the model exhibits a nonequilibrium phase
  transition between the absorbing and endemic phases for $c \geq \lambda^{-1}$, where $c$ is the mean degree and $\lambda$ the average
  infection rate. Interestingly, the
  critical line is independent of the degree distribution but is highly sensitive to the form of the infection-rate distribution.
  We further show that the inverse participation ratio of infection probabilities diverges near the epidemic threshold, indicating that the disease
  may become localized on a small fraction of nodes.
  These results provide a systematic characterization of how network heterogeneities
  shape epidemic spreading on directed contact networks within the quenched mean-field approximation.
\end{abstract}

\maketitle


\section{Introduction}

The susceptible-infected-susceptible (SIS) model on networks provides a minimal yet powerful framework to investigate the interplay between
epidemic spreading and the structure of the underlying contact network \cite{KissBook}. In the SIS model, each node (individual) can be in
either a susceptible or infected state.
A susceptible node $i$ can become infected by a neighbour $j$ at a rate $\lambda_{ij}$, typically assumed to be uniform, $\lambda_{ij} = \lambda$.
An infected node, in turn, becomes healthy at a rate conventionally set to unity. The average fraction of
infected individuals, referred to as the {\it prevalence}, is the natural order-parameter for characterizing the
spread of the epidemic.
In the limit of
an infinitely large number $N$ of individuals, the SIS model may exhibit a nonequilibrium phase transition between an absorbing phase \cite{PastorSatorras2015}, in which
the epidemic dies out, and an endemic phase, characterized by a stationary state with a nonzero prevalence.

A central problem in the study of epidemic spreading is to understand how the network structure
influences the epidemic threshold $\lambda_{\rm c}$ \cite{Chakrabarti2008,Chatterjee2009,Castellano2010,Ferreira2012,Goltsev2012,Li2013,VanMieghem2013,Kwon2013}, which separates the
absorbing phase ($\lambda \leq \lambda_{\rm c}$) from the endemic phase ($\lambda > \lambda_{\rm c}$). The
quenched mean-field (QMF) theory  \cite{Ferreira2012,Mata2013,PastorSatorras2015} provides an effective approximation for the SIS model
by neglecting dynamical correlations between neighboring nodes \cite{Mata2013}, leading to a set of coupled dynamical equations for the single-node
infection probabilities.
Within this framework, the epidemic threshold for a network of size $N$ is given by \cite{Ferreira2012,VanMieghem2013}
\begin{equation}
  \lambda_{\rm c}(N) = 1/\Lambda_1(N),
  \label{hge}
\end{equation}  
where $\Lambda_1(N) > 0$ denotes the leading eigenvalue (spectral radius) of the network adjacency
matrix \cite{Newman2010}.
Beyond its role in QMF theory, Eq. (\ref{hge}) provides a rigorous lower bound for the exact epidemic threshold
on any given network \cite{VanMieghem2013}. The QMF approximation can be improved by incorporating pairwise dynamical
  correlations, leading to the so-called pair quenched mean-field theory \cite{Mata2013}, which yields more accurate
  predictions for the epidemic threshold compared to the individual-based QMF theory.

The spectral properties of the adjacency matrix thus play a fundamental role in the dynamics of
the SIS model.
For undirected networks, the epidemic
threshold follows from well-established analytic results for the leading eigenvalue  $\Lambda_1(N)$ \cite{Krivelevich2003,Chung2003}.
In such networks, the expected value of $\Lambda_1(N)$ typically scales with the maximum degree.
Consequently, for degree distributions with unbounded support, the epidemic threshold vanishes in the limit $N \rightarrow \infty$ \cite{Castellano2010,Ferreira2012}.
This prediction of the QMF theory is
fully consistent with rigorous results for the SIS model \cite{Chatterjee2009}, which prove the absence of an absorbing phase for any nonzero infection rate in undirected
networks.

Another spectral property relevant to the dynamics of the SIS model is the spatial localization of
the leading eigenvector associated with $\Lambda_1(N)$ \cite{Goltsev2012,PastorSatorras2018}.
A strongly localized eigenvector has a finite number of nonzero components as $N \rightarrow \infty$ \cite{AbouChacra1973,Mirlin2000}, whereas
an extended eigenvector is characterized by an extensive number of nonzero components.
Localization effects are expected to be particularly relevant near the epidemic threshold, where the leading eigenvector closely approximates
the stationary endemic state, providing insights into network-based immunization strategies that target
influential nodes \cite{PastorSatorras2018}.
In undirected networks, the localization of the leading eigenvector has been studied through
the inverse participation ratio (IPR) \cite{Goltsev2012,PastorSatorras2016,PastorSatorras2018,Liu2019}, defined in terms of the fourth moment
of the eigenvector components. Numerical results for the IPR indicate that, in finite-size undirected networks, the leading eigenvector
is localized on a vanishing fraction of nodes \cite{Goltsev2012,PastorSatorras2016,PastorSatorras2018}.
In particular, for degree distributions
that decay sufficiently fast, this localization occurs at the hub with the largest degree \cite{Goltsev2012,PastorSatorras2016}.

In contrast to undirected networks, the SIS model on directed contact networks, where the coupling strengths or infection
rates $\lambda_{ij}$ are unidirectional \footnote{In a directed contact network, if node $j$ infects node $i$ with rate $\lambda_{ij}$, the opposite
rate $\lambda_{ji}$ is zero.}, remains poorly understood.
Spreading models on directed networks provide a natural framework for studying the
transmission of computer viruses through email networks \cite{Newman2002,Ebel2002}, the diffusion
of information in social networks \cite{Krueger2024}, and the transmission of diseases between patients
and health care workers \cite{Meyers2006}.
Only a few works have examined how directionality impacts the SIS model \cite{Li2013,Kwon2013}. 
Numerical studies on finite networks with both directed and bidirected edges have shown
that $\lambda_{\rm c}(N)$ increases with the fraction of directed links \cite{Li2013}.
In addition, the SIS model has been studied on directed networks using heterogeneous
mean-field theory \cite{Kwon2013}, which does not fully capture the underlying network structure.
As a result, even basic problems, such as how the directed network structure affects the epidemic threshold, remain unresolved.

In this work, we study the QMF theory of the SIS model on directed networks with arbitrary distributions of degrees and
infection rates. We determine the epidemic threshold $\lambda_{\rm c}$ in the limit $N \rightarrow \infty$, showing that this model undergoes
an absorbing phase transition as a function of the network structure.
Our calculation of $\lambda_{\rm c}$ partially relies on random-matrix analytic results 
for the leading eigenpair of directed weighted networks  \cite{Metz2016,Metz2019,Metz2021}.
In contrast with undirected networks, the leading eigenvalue of directed networks remains finite as $N \rightarrow \infty$ \cite{Metz2016}, even when the degree distribution
has  unbounded support.

We show that the epidemic threshold is determined by the leading eigenvalue only when the network parameters are such that the gap between the leading and the subleading eigenvalue
remains finite as $N \rightarrow \infty$. With the main goal of computing $\lambda_{\rm c}$ for arbitrary combinations of network parameters, we
derive an exact equation for the full distribution of the stationary infection
probabilities in the limit $N \rightarrow \infty$, using the cavity method from spin-glass theory \cite{Mezard2001,Metz2010}. The
numerical solutions of this equation yield both the epidemic threshold and the IPR as functions of the network structure.
We find that $\lambda_{\rm c}$ is independent of the degree distribution, whereas the shape of the infection-rate distribution
strongly influences the epidemic threshold, in particular for large fluctuations in the infection rates.
Furthermore, we show that the IPR diverges near the epidemic threshold as $N \rightarrow \infty$, in agreement
with analytic predictions derived from the moments of the leading eigenvector \cite{Metz2021}. This divergence
arises from a large fraction of nodes with infection probabilities that strongly fluctuate near zero, indicating
that the disease becomes localized on a vanishing fraction of network nodes.

The paper is organized as follows. In the next section, we introduce the SIS model on directed complex networks within the QMF approximation.
In section \ref{cavsec}, we derive the equation for the stationary distribution of infection probabilities using
the cavity method. The results for the epidemic threshold and the IPR as functions of the network parameters
are discussed in section \ref{resultssec}. Finally, we present a summary of our findings and concluding remarks
in section \ref{finalsec}.


\section{The SIS model on directed networks} \label{yuyu}

We consider the SIS model in the quenched mean-field approximation \cite{PastorSatorras2015,Mata2013}. The probability $\rho_i(t)$ ($i=1,\dots,N$) that
a node $i$ is infected at time $t$ evolves according to
\begin{equation}
  \frac{d \rho_i}{d t} = - \rho_i(t) + \left[ 1 - \rho_i(t) \right] \sum_{j=1}^{N} A_{ij} \rho_j(t),
  \label{gtrwe}
\end{equation}  
where the elements $\{ A_{ij} \}_{i,j=1,\dots,N}$ of the $N \times N$
weighted adjacency matrix $\boldsymbol{A}$ have the form $A_{ij} = C_{ij} \lambda_{ij}$, with $C_{ij} \in \{ 0,1 \}$  and $\lambda_{ij} > 0$. The binary variables $\{ C_{ij} \}_{i,j=1,\dots,N}$ specify
the structure of the contact network: if $C_{ij} =1$, there is a directed link $j \rightarrow i$ from node $j$ to $i$, whereas $C_{ij}=0$ otherwise.
The diagonal elements $\{ C_{ii} \}_{i=1,\dots,N}$ are zero.
The indegree $K_i$ (outdegree $L_i$) of node $i$ is given by $K_i = \sum_{j=1}^N C_{ij}$ ($L_i = \sum_{j=1}^N C_{ji}$). We consider
directed random networks generated from the configuration model \cite{Newman2001,Fosdick2018}, where the degree sequences $K_1,\dots,K_N$ and $L_1,\dots,L_N$ are independent and identically
distributed random variables drawn from  $p_{k \ell} = p_{{\rm in},k} p_{{\rm out},\ell}$, with $p_{{\rm in},k}$ and $p_{{\rm out},\ell}$ denoting, respectively, the indegree
and outdegree distributions. The degree sequences fulfill the constraint $\sum_{i=1}^N K_i = \sum_{i=1}^N L_i$, and the average degree $c$ reads
\begin{equation}
c = \sum_{k=0}^{\infty} k \, p_{{\rm in},k} = \sum_{\ell=0}^{\infty} \ell \, p_{{\rm out},\ell}.
\end{equation}  
We will present results for three examples of degree distributions: Poisson, geometric and power-law \cite{Newman2001}. 

The coefficient $\lambda_{ij} > 0$ denotes the directed coupling strength $j \rightarrow i$ or the pairwise infection rate at which node $j$ infects $i$. We assume that
$\{ \lambda_{ij} \}_{i,j=1,\dots,N}$ are independent and identically distributed
random variables sampled from a distribution $P_{\lambda}(x=\lambda_{ij})$ with mean $\lambda$ and variance $\sigma^2$. We will present results for
two choices of $P_{\lambda}(x)$. The first one is the $\Gamma$-distribution
\begin{equation}
  P_{\lambda,{\rm g}}(x) = \frac{\beta^{\alpha}}{\Gamma(\alpha)}  x^{\alpha -1} e^{- \beta x } \Theta(x),
  \label{hjhj}
\end{equation}  
with parameters
\begin{equation}
\alpha = \lambda^2/\sigma^2 \quad \text{and} \quad \beta = \lambda/\sigma^2,
\end{equation}
and with $\Theta(x)$ representing the Heaviside step function.
The second example
is the Pareto distribution with finite variance,
\begin{equation}
P_{\lambda, {\rm p}}(x) = \frac{\gamma x_{0}^{\gamma}}{x^{\gamma +1}} \Theta \left( x - x_0 \right),
\end{equation}  
where 
\begin{equation}
\gamma = 1 + \sqrt{1 + \frac{\lambda^2}{\sigma^2}  } > 2 \quad \text{and} \quad x_0 = \left( \frac{ \sqrt{1 + \frac{\lambda^2}{\sigma^2}  }  }{ 1 + \sqrt{1 + \frac{\lambda^2}{\sigma^2}  }  } \right) \lambda.
\end{equation}  
Note that both distributions are solely parametrized in terms of $\lambda$ and $\sigma$. For $\sigma=0$, the infection rates are uniform ($\lambda_{ij} = \lambda$).

The network ensemble is fully specified
by the distributions $p_{k \ell}$ and $P_{\lambda}(x)$.
For $c > 1$, the network contains a giant strongly
connected component \cite{Kryven2016}, ensuring that the spectrum of $\boldsymbol{A}$ has a continuous component in the limit $N \rightarrow \infty$ \cite{Neri2020}.
This is the interesting regime where the nodes strongly interact with each other and the disease can
eventually infect a finite fraction of individuals.

Our goal is to understand how the structure of the contact network influences
the stationary states of the model. Setting $\frac{d \rho_i}{d t}=0$ in Eq. (\ref{gtrwe}), we obtain
the fixed-point equations
\begin{equation}
  \rho_{i} = \frac{\sum_{j=1}^N A_{ij} \rho_{j}}{1 + \sum_{j=1}^N A_{ij} \rho_{j}  } \quad (i=1,\dots,N).
  \label{fixed}
\end{equation}
To characterize the stationary behaviour and the phase diagram in the limit $N \rightarrow \infty$, we study
the distribution
\begin{equation}
\mathcal{P}(\rho) = \lim_{N \rightarrow \infty} \frac{1}{N} \sum_{i=1}^N \delta (\rho - \rho_{i})
\end{equation}  
of the fixed-point infection probabilities $\rho_{1},\dots,\rho_{N}$. The moments of $\mathcal{P}(\rho)$ read
\begin{equation}
\langle \rho^n \rangle = \lim_{N \rightarrow \infty} \frac{1}{N} \sum_{i=1}^N \rho_{i}^n = \int_{0}^{1} d \rho \mathcal{P}(\rho) \rho^n.
\end{equation}  
In particular, the prevalence $\langle \rho \rangle$ quantifies the average fraction
of infected individuals.

In the next section, we develop an analytic approach, based on the cavity method of spin-glass
theory \cite{Mezard2001,Metz2010}, that yields the following self-consistent equation for the distribution $\mathcal{P}(\rho)$, 
\begin{align}
 & \mathcal{P}(\rho) = \frac{1}{Z} \sum_{k=0}^{\infty} p_{{\rm in},k}  \left( \prod_{j=1}^k \int_{0}^{1}  d \rho_{j} \mathcal{P}(\rho_j)  
   \int_{0}^{\infty}  d x_{j} P_{\lambda}(x_j)   \right) \nonumber \\
   & \times \delta \Bigg[ (1-\rho) \sum_{j=1}^k x_j \rho_j - \rho \Bigg],
   \label{hhjj2}
\end{align}  
where the constant $Z$ ensures that $\int_{0}^{1} d \rho \mathcal{P}(\rho) = 1$. This equation
can be also inferred from the effective dynamics obtained in \cite{Metz2025} using
dynamical mean-field theory.

Equation (\ref{hhjj2}) can be numerically solved through a Monte-Carlo iterative method known as population
dynamics \cite{Mezard2001,MezardBook,Metz2016,Metz2019}. The method is based on the discretization of $\mathcal{P}(\rho)$ in terms of a population with $M$ stochastic
variables $\rho_{1},\dots,\rho_{M}$. In the standard version of the algorithm \cite{Mezard2001,MezardBook,Metz2016}, each iteration
proceeds by randomly selecting a member of the population and updating it 
according to the constraint imposed by the Dirac-$\delta$ in Eq. (\ref{hhjj2}). After a sufficient number of iterations, the population $\rho_{1},\dots,\rho_{M}$ converges
to a stationary profile, yielding an approximate solution of Eq. (\ref{hhjj2}).
In the present context, the core of the algorithm remains unchanged, but we have to include the weight arising from the normalization factor $Z$, which
determines the number of elements updated in parallel at each iteration. The details of this modified version of the algorithm are discussed in \cite{Metz2019}.


\section{The cavity approach for the distribution of infection probabilities}
\label{cavsec}
In this section, we present a first-principles derivation of Eq. (\ref{hhjj2}) using the cavity
  method from spin-glass theory. The dynamic cavity method has been developed in the context of epidemic
  spreading to estimate the probability $\rho_i(t)$ that node $i$ is infected
  at time $t$ \cite{Lokhov2014,Lokhov2015,Shrestha2015,Ortega2022}, leading to important
  applications such as the inference of the origin of an epidemic outbreak \cite{Lokhov2014,Altarelli2014}
  and the design of optimal immunization strategies \cite{Altarelli2014A}. Here, instead of following a
  dynamical approach, we start from the fixed-point equations
  for the infection probabilities $\rho_1,\dots,\rho_N$ in the QMF approximation and compute their empirical distribution using the cavity method.
 We note that this section can be skipped by readers primarily interested in the results obtained by solving Eq. (\ref{hhjj2}).

The variables $\boldsymbol{\rho} = (\rho_1,\dots,\rho_N)^{T}$ solve the equations
\begin{equation}
  F_{N,i}(\boldsymbol{\rho}|\boldsymbol{h}) = - \rho_i + (1-\rho_i) \sum_{j=1}^N A_{ij} \rho_j + h_i = 0,
  \label{huhu33}
\end{equation}  
where $i=1,\dots,N$. We have added auxiliary fields $\boldsymbol{h} =  (h_1,\dots,h_N)^{T}$ to the fixed-point equations, which are useful to identify
the physical meaning of certain quantities. These fields will be set to zero
at the end of the calculation. The joint probability density of $\boldsymbol{\rho}$ can be formally written as \cite{Kurchan1991}
\begin{equation}
  \mathcal{P}_N(\boldsymbol{\rho}|\boldsymbol{h} ) = \dfrac{\prod_{i=1}^N \delta \left[ F_{N,i}(\boldsymbol{\rho}|\boldsymbol{h} )  \right]}{ \int\limits_{0}^{1}
    \left( \prod_{i=1}^N  d \rho_i \right) \prod_{i=1}^N \delta \left[ F_{N,i}(\boldsymbol{\rho} |\boldsymbol{h} )  \right] }.
  \label{hjkl2}
\end{equation}  
Assuming that Eqs. (\ref{huhu33}) admit a unique {\it stable} solution, Eq. (\ref{hjkl2}) concentrates as $N \rightarrow \infty$ on the
joint distribution of infection probabilities
characterizing this stable state. 
Starting from Eq. (\ref{hjkl2}), our purpose is to compute the local marginals $\{ \mathcal{P}_{N,i}(\rho_i) \}_{i=1}^N$ on the network
nodes by using the cavity method \cite{Metz2010}. 
Using the Fourier transform
of the Dirac-$\delta$, we write $\mathcal{P}_N(\boldsymbol{\rho} | \boldsymbol{h} )$ as
\begin{equation}
  \mathcal{P}_N(\boldsymbol{\rho} | \boldsymbol{h}) = \int_{-\infty}^{\infty} \left( \prod_{i=1}^N d \hat{\rho}_{i}  \right) \gamma_N (\boldsymbol{\rho},\hat{\boldsymbol{\rho}}| \boldsymbol{h}  ) ,
  \label{uypo0}
\end{equation}
where
\begin{equation}
  \gamma_N (\boldsymbol{\rho},\hat{\boldsymbol{\rho}}| \boldsymbol{h}) = \left[ Z_N(\boldsymbol{h}) \right]^{-1} \exp{\left[H_N(\boldsymbol{\rho},\hat{\boldsymbol{\rho}}| \boldsymbol{h}  )\right]},
        \label{jiji}
\end{equation}  
with
\begin{equation}
  H_N(\boldsymbol{\rho},\hat{\boldsymbol{\rho}}| \boldsymbol{h}) =  i \sum_{j=1}^N h_j \hat{\rho}_j  - i \sum_{j=1}^N \hat{\rho}_j \rho_j + i \sum_{l,j=1}^N A_{lj} \hat{\rho}_l (1 - \rho_l) \rho_j.
  \label{ggx}
\end{equation}
The constant $Z_N(\boldsymbol{h})$ ensures that $\gamma_N (\boldsymbol{\rho},\hat{\boldsymbol{\rho}}| \boldsymbol{h})$ is normalized.
For $\boldsymbol{h} = 0$, we define $\mathcal{P}_N(\boldsymbol{\rho} |0 ) \equiv \mathcal{P}_N(\boldsymbol{\rho})$ and
$\gamma_N (\boldsymbol{\rho},\hat{\boldsymbol{\rho}}| 0) \equiv \gamma_N (\boldsymbol{\rho},\hat{\boldsymbol{\rho}})$.

From Eq. (\ref{uypo0}), the marginal $\mathcal{P}_{N,i}(\rho_i|h_i)$ is given by
\begin{equation}
  \mathcal{P}_{N,i}(\rho_i | h_i) = \int\limits_{-\infty}^{\infty}  d \hat{\rho}_{i}  \gamma_{N,i} (\rho_i,\hat{\rho}_i | h_i ).
  \label{ija}
\end{equation}
Let us extract the $i$-th term from the sums in Eq. (\ref{ggx}), namely
\begin{align}
  H_N(\boldsymbol{\rho},\hat{\boldsymbol{\rho}} | \boldsymbol{h} ) = i h_i \hat{\rho}_i - i \hat{\rho}_i \rho_i + i \rho_i \sum_{j \in \partial_{i}^{\rm out}} \lambda_{ji} \hat{\rho}_j (1-\rho_j) \nonumber \\
  + i \hat{\rho}_i (1- \rho_i) \sum_{j \in \partial_{i}^{\rm in}} \lambda_{ij} \rho_j +  H_{N-1}^{(i)}(\boldsymbol{\rho},\hat{\boldsymbol{\rho}} | \boldsymbol{h}  ).
  \label{ooii9}
\end{align}  
The function $H_{N-1}^{(i)}(\boldsymbol{\rho},\hat{\boldsymbol{\rho}} | \boldsymbol{h}  )$ is defined on the cavity graph $\mathcal{G}_{N-1}^{(i)}$, which is obtained from the original
graph $\mathcal{G}_{N}$ by removing node $i$, its adjacent edges, and the local field $h_i$. The
symbol $\partial_{i}^{\rm out}$ denotes the set of nodes that receive a directed link from $i$ (the out-neighborhood of $i$), while $\partial_{i}^{\rm in}$
represents the set of nodes that have a directed link pointing to $i$ (the in-neighborhood of $i$). The full neighborhood of $i$ is defined
as  $\partial_i = \partial_{i}^{\rm out} \cup \partial_{i}^{\rm in}$.
By substituting Eq. (\ref{ooii9}) in Eq. (\ref{jiji}) and integrating the resulting expression
over all variables except $(\rho_i,\hat{\rho}_i)$, we find
\begin{align}
 &\gamma_{N,i} (\rho_i,\hat{\rho}_i | h_i ) \sim  e^{i \hat{\rho}_i h_i  -i \hat{\rho}_i \rho_i}   \int\limits_{-\infty}^{\infty}  \Bigg(  \prod_{j \in \partial_i} d \hat{\rho}_{j} \Bigg)
  \int\limits_{0}^{1}  \Bigg(  \prod_{j \in \partial_i} d \rho_{j} \Bigg) \nonumber \\
  & \times  \exp{\Bigg[ i \rho_i \sum_{j \in \partial_{i}^{\rm out}} \lambda_{ji} \hat{\rho}_j (1-\rho_j)  + i \hat{\rho}_i (1- \rho_i) \sum_{j \in \partial_{i}^{\rm in}} \lambda_{ij} \rho_j  \Bigg] } \nonumber \\
  & \times \gamma_{N-1,\partial_i}^{(i)} (\rho_{\partial_i},\hat{\rho}_{\partial_i} | h_{\partial_i } ),
  \label{jjbbnn}
\end{align}  
where $\gamma_{N-1,\partial_i}^{(i)} (\rho_{\partial_i},\hat{\rho}_{\partial_i} | h_{\partial_i } )$ is the joint distribution defined
on the neighbourhood $\partial_i$ within the cavity graph $\mathcal{G}_{N-1}^{(i)}$.
We have omitted the normalization
constant in Eq. (\ref{jjbbnn}), as this quantity can be fixed at the end of the calculation.

At this point, we invoke the main assumption of the cavity method on sparse random graphs \cite{Mezard2001,Metz2010}.
Since networks generated from the configuration model become locally tree-like for $N \gg 1$, the function $\gamma_{N-1,\partial_i}^{(i)} (\rho_{\partial_i},\hat{\rho}_{\partial_i} |  h_{\partial_i } )$
factorizes as follows
\begin{equation}
\gamma_{N-1,\partial_i}^{(i)} (\rho_{\partial_i},\hat{\rho}_{\partial_i} |  h_{\partial_i }  ) = \prod_{j \in \partial_i} \gamma_{N-1,j}^{(i)} (\rho_{j},\hat{\rho}_{j}|h_j ).
\end{equation}  
Inserting the above expression in Eq. (\ref{jjbbnn}) and using Eq. (\ref{ija}), we find
\begin{align}
  &\gamma_{N,i} (\rho_i,\hat{\rho}_i | h_i ) \sim  e^{i \hat{\rho}_i h_i  -i \hat{\rho}_i \rho_i} \nonumber \\
  &\times \prod_{j \in \partial_{i}^{\rm in}}  \int\limits_{0}^{1} d \rho_{j}
  \, e^{  i \hat{\rho}_i (1- \rho_i)  \lambda_{ij} \rho_j    } \, \mathcal{P}_{N-1,j}^{(i)} (\rho_{j} | h_j) \nonumber \\
  &\times \prod_{j \in \partial_{i}^{\rm out}}  \int\limits_{0}^{1} d \rho_{j} \int\limits_{-\infty}^{\infty} d \hat{\rho}_{j}
  \, e^{  i \hat{\rho}_j (1- \rho_j)  \lambda_{ji} \rho_i    } \, \gamma_{N-1,j}^{(i)} (\rho_{j},\hat{\rho}_{j} | h_j). \label{popo9}
\end{align}  
In the terminology of the message-passing algorithm \cite{MezardBook}, $\gamma_{N-1,j}^{(i)}(\rho_{j},\hat{\rho}_{j} | h_j)$ denotes the message that
propagates from node $j$ to $i$ along the directed edge $j \rightarrow i$.
Since a node $i$ can be infected only by its in-neighbourhood $\partial_{i}^{\rm in}$, the term containing the messages $\gamma_{N-1,j}^{(i)} (\rho_{j},\hat{\rho}_{j} | h_j)$ originated
in the out-neighbourhood $j \in \partial_{i}^{\rm out}$ of $i$ should be irrelevant and contribute with a constant in Eq. (\ref{popo9}).

Let us now demonstrate this property. By repeating the same reasoning that led to Eq. (\ref{popo9}), we can derive the cavity equations   for $\gamma_{N-1,j}^{(i)} (\rho_{j},\hat{\rho}_{j} | h_j)$.
For $i \in  \partial_{j}^{\rm in}$, we obtain
\begin{align}
  &\gamma_{N-1,j}^{(i)} (\rho_j,\hat{\rho}_j | h_j ) \sim  e^{i \hat{\rho}_j h_j  -i \hat{\rho}_j \rho_j} \nonumber \\
  &\times \prod_{l \in \partial_{j}^{\rm in} \setminus i} \, \int\limits_{0}^{1} d \rho_{l}
  \, e^{  i \hat{\rho}_j (1- \rho_j)  \lambda_{jl} \rho_l    } \, \mathcal{P}_{N-1,l}^{(j)} (\rho_{l} | h_l) \nonumber \\
  &\times \prod_{l \in \partial_{j}^{\rm out}} \int\limits_{-\infty}^{\infty} d \hat{\rho}_{l} \int\limits_{0}^{1} d \rho_{l}
  \, e^{  i \hat{\rho}_l (1- \rho_l)  \lambda_{lj} \rho_j    } \, \gamma_{N-1,l}^{(j)} (\rho_{l},\hat{\rho}_{l} | h_l), \label{popo10}
\end{align}  
while for $i \in  \partial_{j}^{\rm out}$ we get
\begin{align}
  &\gamma_{N-1,j}^{(i)} (\rho_j,\hat{\rho}_j | h_j ) \sim  e^{i \hat{\rho}_j h_j  -i \hat{\rho}_j \rho_j} \nonumber \\
  &\times \prod_{l \in \partial_{j}^{\rm in}} \, \int\limits_{0}^{1} d \rho_{l}
  \, e^{  i \hat{\rho}_j (1- \rho_j)  \lambda_{jl} \rho_l    } \, \mathcal{P}_{N-1,l}^{(j)} (\rho_{l} | h_l) \nonumber \\
  &\times \prod_{l \in \partial_{j}^{\rm out}  \setminus i} \int\limits_{-\infty}^{\infty} d \hat{\rho}_{l} \int\limits_{0}^{1} d \rho_{l}
  \, e^{  i \hat{\rho}_l (1- \rho_l)  \lambda_{lj} \rho_j    } \, \gamma_{N-1,l}^{(j)} (\rho_{l},\hat{\rho}_{l} | h_l), \label{popo34}
\end{align}  
where $\partial_{j}^{\rm in} \setminus i$ ($\partial_{j}^{\rm out} \setminus i$)  represents the set of nodes in the in-neighbourhood (out-neighbourhood) of $j$, except for node $i$.
Let us focus on $\gamma_{N-1,j}^{(i)} (\rho_j,\hat{\rho}_j | h_j )$ with $i \in \partial_{j}^{\rm in}$, which is precisely the quantity appearing in the right hand
side of Eq. (\ref{popo9}).
By setting all auxiliary fields to zero in Eq. (\ref{popo10}), except for $h_j$, we obtain the interesting identity
\begin{equation}
\gamma_{N-1,j}^{(i)} (\rho_j,\hat{\rho}_j | h_j ) = e^{i \hat{\rho}_j h_j} \gamma_{N-1,j}^{(i)} (\rho_j,\hat{\rho}_j),
\end{equation}  
with $\gamma_{N-1,j}^{(i)} (\rho_j,\hat{\rho}_j ) \equiv \gamma_{N-1,j}^{(i)} (\rho_j,\hat{\rho}_j | 0 )$. Hence, setting all external fields to zero
in Eq. (\ref{popo9}) and using the above identity, we find
\begin{align}
  &\gamma_{N,i} (\rho_i,\hat{\rho}_i) \sim  e^{-i \hat{\rho}_i \rho_i} \prod_{j \in \partial_{i}^{\rm in}}  \int\limits_{0}^{1} d \rho_{j}
  \, e^{  i \hat{\rho}_i (1- \rho_i)  \lambda_{ij} \rho_j    } \, \mathcal{P}_{N-1,j}^{(i)} (\rho_{j}) \nonumber \\
  &\times \prod_{j \in \partial_{i}^{\rm out}}  \int\limits_{0}^{1} d \rho_{j} \int\limits_{-\infty}^{\infty} d \hat{\rho}_{j}
  \, \gamma_{N-1,j}^{(i)} (\rho_{j},\hat{\rho}_{j} |(1- \rho_j)  \lambda_{ji} \rho_i     ).
  \label{popo78}
\end{align}  
The final step of this calculation is to understand the physical meaning
of $\gamma_{N-1,j}^{(i)} (\rho_{j},\hat{\rho}_{j} |(1- \rho_j)  \lambda_{ji} \rho_i )$.
This can be achieved by extracting the $j$-th term 
from the function $H_{N-1}^{(i)}(\boldsymbol{\rho},\hat{\boldsymbol{\rho}} | \boldsymbol{h} )$ on the cavity
graph $\mathcal{G}_{N-1}^{(i)}$, 
\begin{align}
  H_{N-1}^{(i)} (\boldsymbol{\rho},\hat{\boldsymbol{\rho}} | \boldsymbol{h} ) = i h_j \hat{\rho}_j - i \hat{\rho}_j \rho_j + i \rho_j \sum_{l \in \partial_{j}^{\rm out}} \lambda_{lj} \hat{\rho}_l (1-\rho_l) \nonumber \\
  + i \hat{\rho}_j (1- \rho_j) \sum_{l \in \partial_{j}^{\rm in} \setminus i} \lambda_{jl} \rho_l +  H_{N-2}^{(i,j)}(\boldsymbol{\rho},\hat{\boldsymbol{\rho}} | \boldsymbol{h}  ),
  \label{ooii10}
\end{align}  
where $i \in \partial_{j}^{\rm in}$.
If we choose $h_j = (1- \rho_j)  \lambda_{ji} \rho_i$ in the above equation, we restore the neighbourhood $\partial_j$ of $j$ as in the original graph.
In other words, by setting the external field to $h_j = (1- \rho_j)  \lambda_{ji} \rho_i$, we add back the interaction term between $j$ and $i$, which has been deleted through
the removal of $i$. Hence, for $N \gg 1$, the following relation holds 
\begin{equation}
\gamma_{N-1,j}^{(i)} (\rho_{j},\hat{\rho}_{j} |(1- \rho_j)  \lambda_{ji} \rho_i ) = \gamma_{N,j} (\rho_{j},\hat{\rho}_{j} ).
\end{equation}  
Inserting the above relation in Eq. (\ref{popo78}) and using the normalization of $\gamma_{N}(\rho_{j},\hat{\rho}_{j} )$, we obtain
\begin{equation}
  \gamma_{N,i} (\rho_i,\hat{\rho}_i ) \sim  e^{-i \hat{\rho}_i \rho_i} \prod_{j \in \partial_{i}^{\rm in}}  \int\limits_{0}^{1} d \rho_{j}
  \, e^{  i \hat{\rho}_i (1- \rho_i)  \lambda_{ij} \rho_j    } \, \mathcal{P}_{N-1,j}^{(i)} (\rho_{j}),
  \label{popo80}
\end{equation}  
as expected.

Finally, by integrating Eq. (\ref{popo80}) over $\hat{\rho}_{i}$, we find 
\begin{align}
  \mathcal{P}_{N,i}(\rho) &= \frac{1}{Z_{N,i}}  \int\limits_{0}^{1} \Bigg( \prod_{j \in \partial_{i}^{\rm in}} d \rho_{j} \mathcal{P}_{N-1,j}^{(i)}(\rho_j) \Bigg) \nonumber \\
  &\times \delta \Bigg[ (1-\rho) \sum_{j \in \partial_{i}^{\rm in}} \lambda_{ij} \rho_j - \rho     \Bigg],
  \label{dodo}
\end{align}  
where $Z_{N,i}$ ensures that $\mathcal{P}_{N,i}(\rho)$ is normalized. Hence, for large $N$, $\mathcal{P}_{N,i}(\rho)$ follows from
the local marginals $\{ \mathcal{P}_{N-1,j}^{(i)}(\rho_j) \}_{j \in \partial_{i}^{\rm in}}$ on $\mathcal{G}_{N-1}^{(i)}$.
To determine the cavity marginals  $\mathcal{P}_{N-1,j}^{(i)}(\rho_j)$, we start from Eq. (\ref{popo34}) and follow exactly the same reasoning as
explained above, which leads to the cavity equations
\begin{align}
  \mathcal{P}_{N-1,i}^{(l)}(\rho) &= \frac{1}{Z_{N-1,i}^{(l)}}  \int\limits_{0}^{1} \Bigg( \prod_{j \in \partial_{i}^{\rm in}} d \rho_{j} \mathcal{P}_{N-1,j}^{(i)}(\rho_j) \Bigg) \nonumber \\
  &\times \delta \Bigg[ (1-\rho) \sum_{j \in \partial_{i}^{\rm in}} \lambda_{ij} \rho_j - \rho     \Bigg],
  \label{htdew}
\end{align}  
with $l \in \partial_{i}^{\rm out}$. The constant $Z_{N-1,i}^{(l)}$ normalizes  $\mathcal{P}_{N-1,i}^{(l)}(\rho)$.
The local marginals $\mathcal{P}_{N,i}(\rho)$ and $\mathcal{P}_{N-1,i}^{(l)}(\rho)$ fulfill the same
  equation on a single graph instance because the quantity
  $\sum_{j \in \partial_{i}^{\rm in}} \lambda_{ij} \rho_j$ is solely determined by the in-neighbourhood $\partial_{i}^{\rm in}$, while $\mathcal{P}_{N-1,i}^{(l)}(\rho)$ is
  obtained by removing a node $l$ from the out-neighbourhood. This is different, for instance, from the cavity approach
  in undirected networks, where there is no distinction between in-neighbourhood and out-neighbourhood, and the removal of a node influences
  the states of all its neighbours.

The solutions of Eqs. (\ref{dodo}) and (\ref{htdew}) provide
accurate approximations for the local marginals $\{ \mathcal{P}_{N,i}(\rho) \}_{i=1,\dots,N}$ on single
network instances with large $N$ and a locally tree-like structure.
Equation (\ref{htdew}) becomes asymptotically exact as $N \rightarrow \infty$. In this limit, we introduce the ensemble-averaged quantities
\begin{equation}
\mathcal{P}(\rho) = \lim_{N \rightarrow \infty} \frac{1}{N} \sum_{j=1}^N  \mathcal{P}_{N,j}(\rho)
\end{equation}  
and
\begin{equation}
\mathcal{P}_{\rm edg} (\rho) = \lim_{N \rightarrow \infty} \frac{1}{N c} \sum_{j=1}^N \sum_{l \in \partial_{j}^{\rm out} }  \mathcal{P}_{N-1,j}^{(l)}(\rho).
\end{equation}  
Assuming that in Eq. (\ref{dodo}) the numerator and denominator converge
independently to their ensemble-averaged values as $N \rightarrow \infty$, we
conclude that $\mathcal{P}(\rho)$ is determined by
\begin{align}
  \mathcal{P}(\rho) &= \frac{1}{Z} \sum_{k,\ell=0}^{\infty} p_{k \ell}  \Bigg( \prod_{j =1}^k  \int\limits_{0}^{1} d \rho_{j} \mathcal{P}_{\rm edg} (\rho_j)
\int\limits_{0}^{\infty} d x_j P_{\lambda}(x_j) \Bigg) 
   \nonumber \\
   &\times  \delta \Bigg[ (1-\rho) \sum_{j =1}^k x_{j} \rho_j - \rho     \Bigg].
   \label{uy1}
\end{align}  
Analogously, $\mathcal{P}_{\rm edg} (\rho)$ fulfills the self-consistent equation
\begin{align}
  \mathcal{P}_{\rm edg} (\rho) &= \frac{1}{Z_{\rm edg}} \sum_{k,\ell=0}^{\infty} \frac{\ell p_{k \ell}  }{c}
 \Bigg( \prod_{j =1}^k  \int\limits_{0}^{1} d \rho_{j} \mathcal{P}_{\rm edg} (\rho_j) \Bigg)   
   \nonumber \\
   &\times  \Bigg( \prod_{j =1}^k  \int\limits_{0}^{\infty} d x_j P_{\lambda}(x_j) \Bigg)  \delta \Bigg[ (1-\rho) \sum_{j =1}^k x_{j} \rho_j - \rho     \Bigg].
    \label{uy2}
\end{align}  
The constants $Z$ and $Z_{\rm edg}$ denote the corresponding normalization factors. Equations (\ref{uy1}) and (\ref{uy2}) are valid for networks with an
arbitrary joint distribution $p_{k \ell}$ of indegrees and outdegrees. Substituting $p_{k \ell} = p_{{\rm in},k} p_{{\rm out},\ell}$ in Eq. (\ref{uy2}) and summing
  over $l$, we conclude that $\mathcal{P}(\rho)=\mathcal{P}_{\rm edg} (\rho)$, recovering Eq. (\ref{hhjj2}). 


\section{Results}
\label{resultssec}

In this section, we determine the phase diagram of the SIS model in the limit $N \rightarrow \infty$ by combining analytic results
from random matrix theory \cite{Neri2016,Metz2019,Metz2021} with numerical solutions of Eq. (\ref{hhjj2}).

\subsection{Linear stability analysis}

First, we perform a linear stability analysis of the disease-free fixed-point $\rho_{i} = 0$ ($i=1,\dots,N$) that characterizes the absorbing phase. The linearized
form of Eq. (\ref{gtrwe}) is given by
\begin{equation}
  \frac{d \boldsymbol{\rho}}{d t} = \left(\boldsymbol{A} - \boldsymbol{I} \right) \boldsymbol{\rho}(t),
  \label{jhdt}
\end{equation}  
where $\boldsymbol{\rho}(t) = (\rho_1(t),\dots,\rho_N(t))^{T}$, and $\boldsymbol{I}$ is the $N \times N$ identity matrix. The solution of the above equation
determines whether perturbations of the trivial fixed-point decay to zero or grow in time.
By introducing the right $\{ \boldsymbol{R}_{\alpha} \}_{\alpha=1,\dots,N}$ and left $\{ \boldsymbol{L}_{\alpha} \}_{\alpha=1,\dots,N}$
eigenvectors of the asymmetric matrix $\boldsymbol{A}$,
\begin{equation}
 \boldsymbol{A} \boldsymbol{R}_{\alpha} = \Lambda_{\alpha} \boldsymbol{R}_{\alpha} \quad \boldsymbol{L}_{\alpha}  \boldsymbol{A} = \Lambda_{\alpha} \boldsymbol{L}_{\alpha},
\end{equation}
the solution of Eq. (\ref{jhdt}) reads
\begin{equation}
  \boldsymbol{\rho}(t) = \sum_{\alpha=1}^N  \left[ \boldsymbol{L}_{\alpha} \, \boldsymbol{\rho}(0) \right]  e^{(\Lambda_{\alpha} - 1 ) t } \boldsymbol{R}_{\alpha},
  \label{uyuy6}
\end{equation}  
where $\Lambda_{1},\dots,\Lambda_{N}$ are the eigenvalues of $\boldsymbol{A}$. By ordering the eigenvalues
according to their real parts,  ${\rm Re} \Lambda_1 \geq {\rm Re} \Lambda_2 \geq \dots \geq  {\rm Re} \Lambda_N$, we conclude
that $|\boldsymbol{\rho}(t)|$ decays to zero for $t \rightarrow \infty$  provided
\begin{equation}
   {\rm Re} \Lambda_1 < 1.
  \label{ytws}
\end{equation}  
The above condition determines the linear stability of the disease-free fixed-point.

In a series of previous works \cite{Neri2016,Metz2019,Metz2021}, the leading eigenvalue $\Lambda_1$ and the statistical properties of the corresponding
right eigenvector $\boldsymbol{R}_{1}$ of sparse directed networks have been computed in the limit $N \rightarrow \infty$. For $c > 1$, the spectral
density of $\boldsymbol{A}$ has a continuous
component \cite{Neri2020} and the spectral gap $|\Lambda_1-\Lambda_2|$ is finite if $c > c_{\rm gap}$, where
\begin{equation}
c_{\rm gap} = 1 + \sigma^2/\lambda^2.
\end{equation}  
For $c > c_{\rm gap}$, $\Lambda_1 = c \lambda$ is an outlier \cite{Neri2016}, and the trivial solution is
stable if $c < \lambda^{-1}$.
For $c \leq  c_{\rm gap}$, the spectral gap is zero, $\Lambda_1 = \sqrt{c \left(\sigma^2 + \lambda^2 \right)} \in \mathbb{R}$ belongs to the boundary of
the continuous spectrum, and the trivial fixed-point is stable for
\begin{equation}
  c  < \left( \sigma^2 + \lambda^2  \right)^{-1}.
  \label{hu2}
\end{equation}  
These results for the stability analysis hold for $c > 1$. The stability thresholds
are universal, as they are independent of higher moments
of the distributions $p_{k \ell}$ and $P_{\lambda}(x)$ characterizing the network structure.

As soon as the trivial fixed-point becomes unstable, the leading eigenvector $\boldsymbol{R}_{1}$ plays a crucial
role for the dynamics. The  moments of the real-valued eigenvector components $\{ R_{1,i} \}_{i=1}^N$
are defined as
\begin{equation}
\langle R_{1}^{n} \rangle = \int\limits_{-\infty}^{\infty} dr \, p_{R}(r) r^n,
\end{equation}  
where the distribution $p_R(r)$ reads
\begin{equation}
p_R(r) = \lim_{N \rightarrow \infty} \frac{1}{N} \sum\limits_{i=1}^N \delta (r - R_{1,i} ).
\end{equation}  
The moments of $p_R(r)$ characterize the fluctuations of the eigenvector components, allowing to study
localization phenomena. In reference \cite{Metz2021}, the first moments of $p_R(r)$ have been analytically computed
for sparse directed networks, unveiling a localization transition as a function of the network parameters.

\subsection{Phase diagram}

\begin{figure}[H]
  \begin{center}
    \includegraphics[scale=0.24]{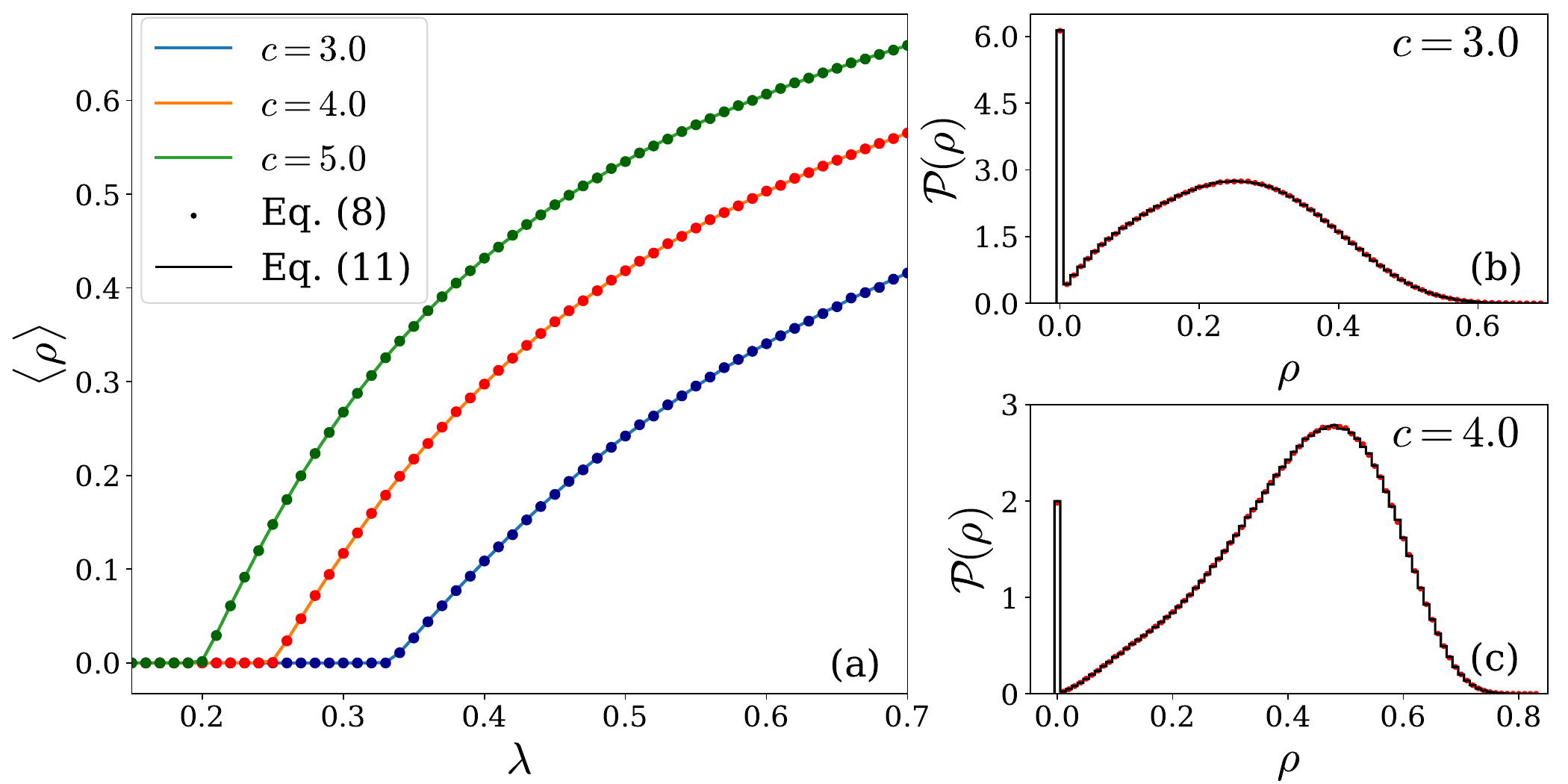} 
    \caption{Comparison between the solutions of Eq. (\ref{hhjj2}) (solid lines) with the fixed-point solutions
      of Eq. (\ref{fixed}) (symbols) for directed networks with a Poisson degree distribution and a $\Gamma$-distribution of infection rates or coupling strengths.
      The fixed-point solutions are derived from an ensemble of $50$ networks with $N=10^5$ nodes, while the population dynamics
      results are obtained from $5$ independent runs of the algorithm with $M=10^6$ stochastic variables. (a) The prevalence $\langle \rho \rangle$ as
      a function of the mean infection rate $\lambda$ for standard
      deviation $\sigma=0.2$ of the infection rates. (b) and (c): distribution $\mathcal{P}(\rho)$ of the infection
      probabilities for $\lambda=1/2$, $\sigma=0.2$, and two
      different $c$.
}
\label{compare}
\end{center}
\end{figure}

Before discussing how the epidemic threshold depends on the network parameters, we compare  the solutions
of Eq. (\ref{hhjj2}), valid for $N \rightarrow \infty$,  with results obtained from the fixed-point Eq. (\ref{fixed}) on finite-sized networks. Figure \ref{compare}
shows numerical results for the prevalence $\langle \rho \rangle$ and the full distribution $\mathcal{P}(\rho)$. In both
cases, the agreement between the solutions of Eq. (\ref{hhjj2})
and those obtained from Eq. (\ref{fixed}) is excellent.
The distribution $\mathcal{P}(\rho)$ features a Dirac-$\delta$ at $\rho=0$, reflecting a finite fraction of nodes
with zero indegree.
While both approaches rely on numerical computations, an important advantage
of Eq. (\ref{hhjj2}) over Eq. (\ref{fixed}) is that the former does not require the use of sophisticated algorithms to sample networks 
from the configuration model, since it depends on the network structure only through $p_{{\rm in},k}$ and $P_{\lambda}(x)$.

\begin{figure}[H]
  \begin{center}
    \includegraphics[scale=0.28]{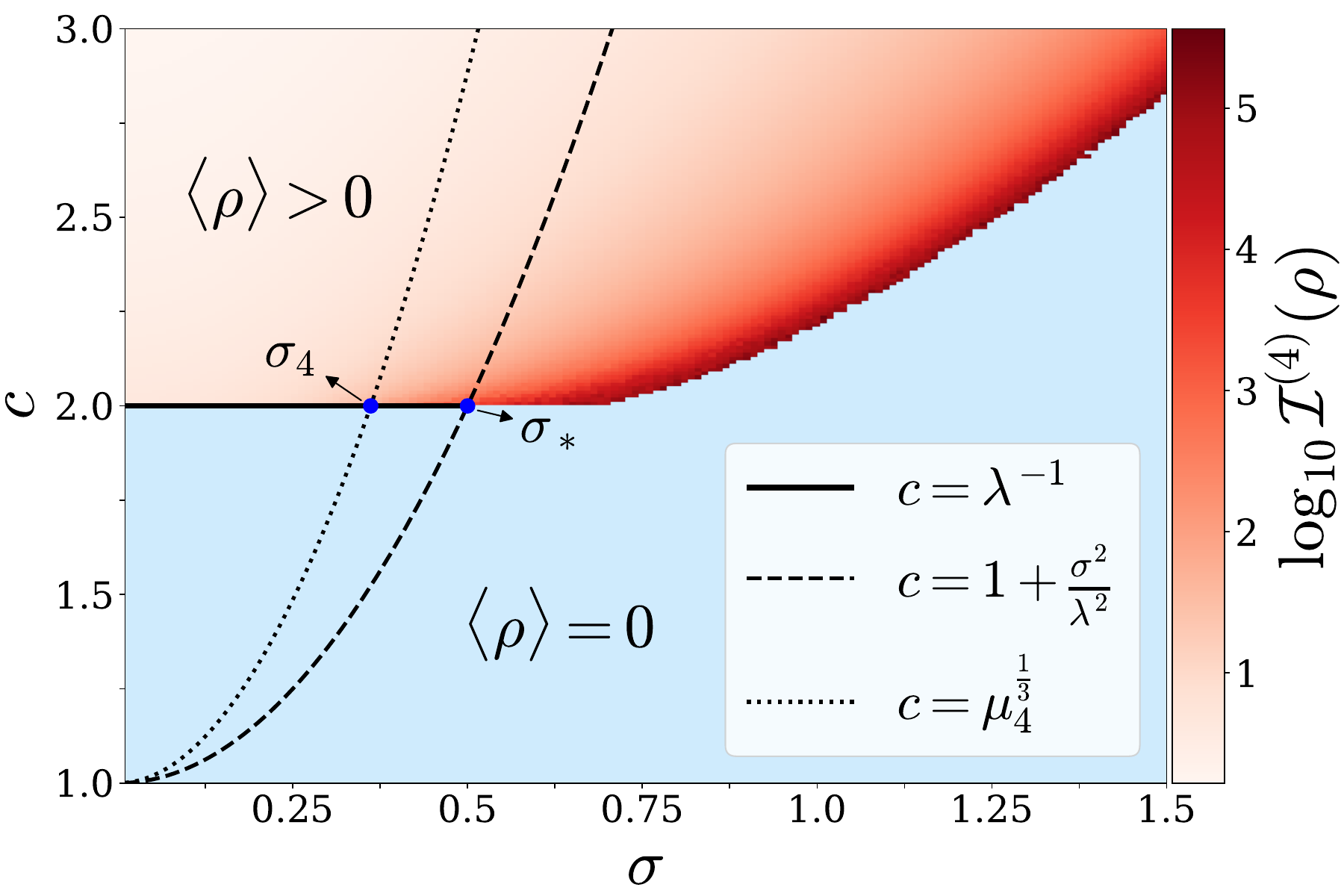} 
    \caption{Phase diagram of the SIS model on  directed networks in terms of the mean degree $c$ and the standard deviation $\sigma$ of
      the infection rates or coupling strengths (the
      mean infection rate is $\lambda=1/2$).
      The indegrees follow a Poisson distribution, while the infection rates follow
      a $\Gamma$-distribution. The model exhibits an endemic phase ($\langle \rho \rangle >0$) and
      an absorbing phase ($\langle \rho \rangle=0$). 
       The standard deviations at the dots are, respectively, $\sigma_{*} = \sqrt{\lambda (1 - \lambda)  }$ and $\sigma_{4} \simeq 0.36$. For $\sigma > \sigma_{*}$, the critical line
       is obtained by solving Eq. (\ref{hhjj2}) using the population dynamics algorithm with $M= 10^6$ stochastic variables. The colour scale
       shows the inverse participation, Eq. (\ref{huhu55}), which quantifies the spatial fluctuations of the infection probabilities. The fourth
       moment $\mu_{4}$ of the infection rates is defined in Eq. (\ref{hhjj}).
}
\label{phasediagram}
\end{center}
\end{figure}

In figure \ref{phasediagram}, we present the phase diagram of the SIS model on directed networks in terms of $(\sigma,c)$. The
model exhibits an absorbing phase, where $\mathcal{P}(\rho) = \delta(\rho)$, and
an endemic phase, characterized by a stationary distribution $\mathcal{P}(\rho)$ with nonzero prevalence $\langle \rho \rangle$.
The average degree $c = \lambda^{-1}$ above which the epidemic spreads to a finite fraction of the population is determined solely by the
mean infection rate and is independent of other network properties. If $(\sigma,c)$ lies above the dashed
line in the phase diagram, the spectral gap $|\Lambda_1-\Lambda_2|$ is finite; otherwise, $|\Lambda_1-\Lambda_2| = 0$.
For $\sigma < \sigma_{*}$, the transition line that delimits the absorbing phase follows from the linear stability analysis of the disease-free fixed-point.
For $\sigma > \sigma_{*}$, the transition line is obtained by numerically solving Eq. (\ref{hhjj2}) and monitoring the prevalence $\langle \rho \rangle$.
The colour scale in figure \ref{phasediagram} quantifies the inverse participation ratio of the infection
probabilities, which will be discussed in the next subsection.

The linear stability analysis identifies the leading eigenvector $\boldsymbol{R}_1$ as responsible for destabilizing
the absorbing phase. By the Perron-Frobenius theorem \cite{Horn2012}, the components
of $\{ R_{1,i} \}_{i=1}^N$ are non-negative.
Above the dashed line in figure \ref{phasediagram}, $\boldsymbol{R}_1$ is associated with an outlier
eigenvalue, characterized by $\langle R_1 \rangle > 0$ \cite{Metz2021}. This ``ferromagnetic'' mode triggers the onset of the endemic
phase. Below the dashed line, the leading eigenvalue lies
at the boundary of the continuous spectrum, where $\langle R_1 \rangle = 0$ \cite{Metz2021}. Combined
with the constraint $R_{1,i} \geq 0$, this suggests that $R_{1,i} = 0$ with
probability one in the limit $N \rightarrow \infty$.
This mode is unable to destabilize the absorbing state
$\mathcal{P}(\rho) = \delta(\rho)$, which explains the absence of a transition to the  endemic phase for $c < \lambda^{-1}$, in contrast
with the prediction of Eq. (\ref{hu2}). Hence, the phase transition for $\sigma > \sigma_{*}$ in figure \ref{phasediagram} is not governed by the
leading eigenpair of the contact network.

\begin{figure}[H]
  \begin{center}
    \includegraphics[scale=0.37]{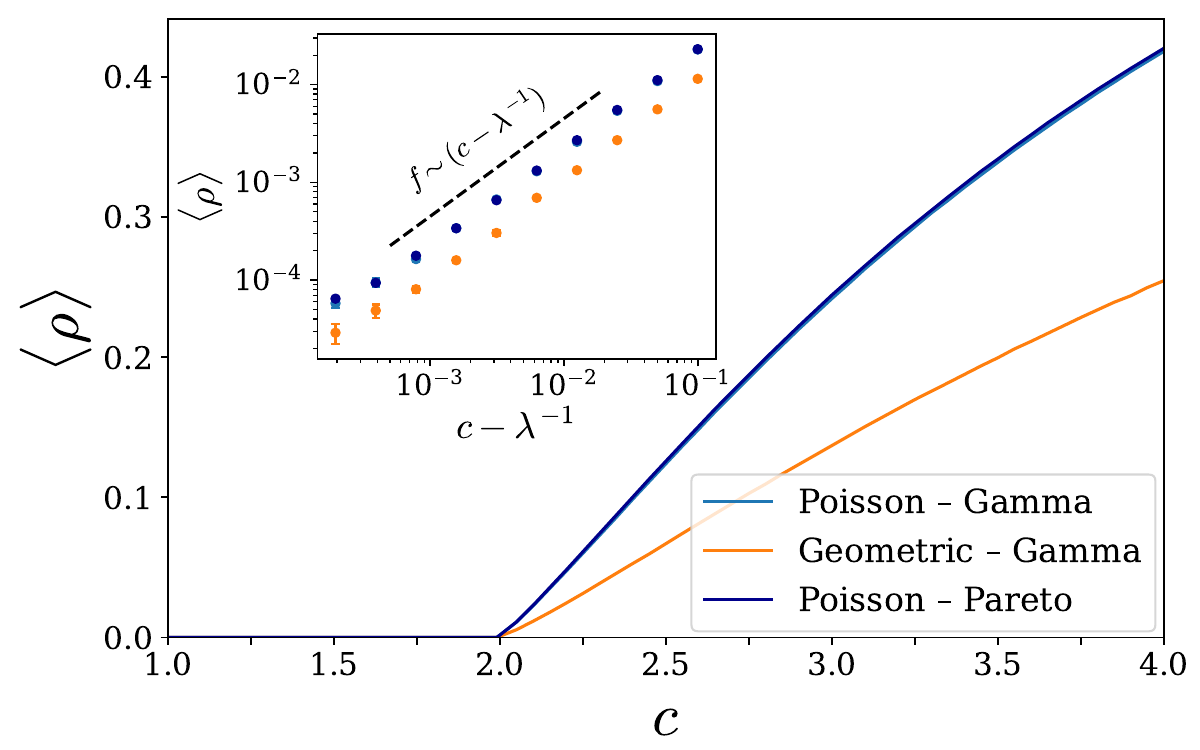} 
    \caption{Prevalence $\langle \rho \rangle$ as a function of the mean degree $c$ for different distributions of infection rates
      and indegrees in the regime $\sigma < \sigma_{*}$ (see the main text).
      The infection rates or coupling strengths have mean $\lambda = 1/2$ and standard deviation $\sigma = 0.2$.
      The results are obtained by solving Eq. (\ref{hhjj2}) using the population dynamics algorithm with $M= 10^5$ stochastic variables.
      The inset shows the prevalence near $c=\lambda^{-1}$ in logarithmic scale. The colours in the inset
      correspond to the same distributions as in the main panel.
}
\label{transition1}
\end{center}
\end{figure}
\begin{figure}[H]
  \begin{center}
    \includegraphics[scale=0.37]{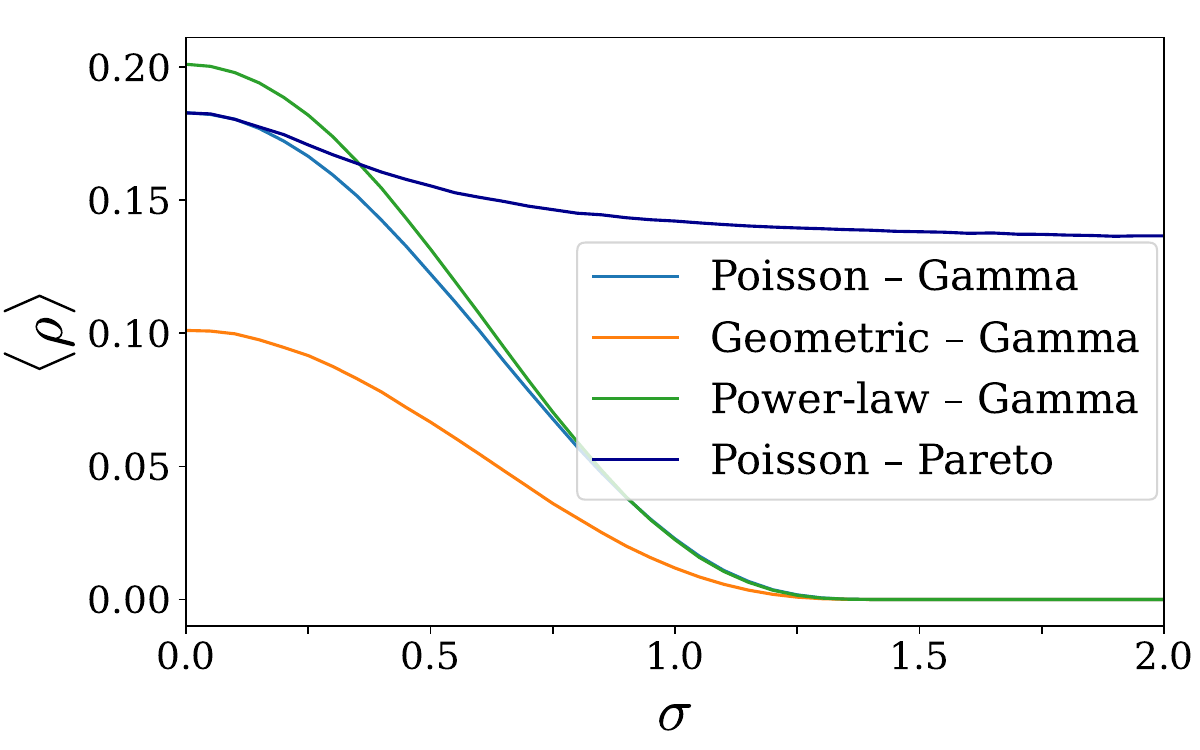} 
    \caption{Prevalence $\langle \rho \rangle$ as a function of the standard deviation $\sigma$ of the infection rates for different distributions of
      indegrees and infection rates in the regime $\sigma > \sigma_{*}$ (see the main text).
      The average indegree is $c \simeq 2.7$, and the infection rates have mean $\lambda = 1/2$. For power-law distributed indegrees, the smallest
      indegree is $k_{\rm min}=2$.
      The results are obtained by solving Eq. (\ref{hhjj2}) using the population dynamics algorithm with $M=10^6$ stochastic variables.     
}
\label{transition2}
\end{center}
\end{figure}

We now discuss the role of network heterogeneities on the prevalence $\langle \rho \rangle$ and the epidemic threshold.
Figure \ref{transition1} shows $\langle \rho \rangle$ as a function of $c$ for different
 distributions $p_{{\rm in},k}$ and $P_{\lambda}(x=\lambda_{ij})$ in the regime $\sigma < \sigma_{*}$. The
prevalence vanishes as $\langle \rho \rangle \simeq c - \lambda^{-1}$ for $0 < c - \lambda^{-1} \ll 1 $, consistent with the mean-field critical exponent
of directed percolation \cite{Odor2004}. The epidemic threshold as well as the critical behaviour of  $\langle \rho \rangle$ are both
independent of the distributions $p_{{\rm in},k}$ and $P_{\lambda}(x)$, confirming the universality of the transition at $c=\lambda^{-1}$.

In the regime $\sigma > \sigma_{*}$, the stationary behaviour becomes highly sensitive to
the shape of the distribution $P_{\lambda}(x)$ of infection rates.
As shown in figure \ref{transition2}, when $\lambda_{ij}$ follows a $\Gamma$-distribution, the prevalence
drops to zero at sufficiently large $\sigma$, due to the large fraction of small infection rates caused by the divergence of $P_{\lambda,{\rm g}}(x=0)$ (see Eq. (\ref{hjhj})).
In contrast, when $\lambda_{ij}$ follows a Pareto distribution, in which the smallest infection rate is $x_0 > 0$, the endemic state persists
even for strong fluctuations of $\lambda_{ij}$, and $\langle \rho \rangle$ saturates at a finite value.
Taken together, figures \ref{transition1} and \ref{transition2} show that the epidemic
threshold is independent of the indegree distribution $p_{{\rm in},k}$.


\subsection{Localization of epidemic spreading}

In this section we present results for higher moments of $\mathcal{P}(\rho)$, which characterize
the spatial fluctuations of the infection probabilities and the localization of the epidemics.
We also discuss the connection between the moments of $\mathcal{P}(\rho)$ and those of the distribution of the leading
eigenvector $\boldsymbol{R}_1$.

The inverse participation ratio (IPR) is a standard probe of spatial localization in disordered systems \cite{Mirlin2000,Goltsev2012,Metz2021}. 
Following \cite{Metz2021}, the IPR of the fixed-point vector $\boldsymbol{\rho} = (\rho_{1},\dots,\rho_{N})^{T}$ is defined as follows
\begin{equation}
\mathcal{I}_N^{(4)} (\boldsymbol{\rho}) = \dfrac{N \sum_{i=1}^N \rho_{i}^4}{\left( \sum_{i=1}^N \rho_{i}^2 \right)^{2} },
\end{equation}  
while the dimensionless second moment reads
\begin{equation}
\mathcal{I}_N^{(2)} (\boldsymbol{\rho}) = \dfrac{N \sum_{i=1}^N \rho_{i}^2}{\left( \sum_{i=1}^N \rho_{i} \right)^{2} }.
\end{equation}  
If the network has a finite number of nodes with nonzero infection probabilities, $\mathcal{I}_N^{(4)}(\boldsymbol{\rho})$ and $\mathcal{I}_N^{(2)}(\boldsymbol{\rho})$ both scale linearly
with $N$ and the vector $\boldsymbol{\rho}$ is localized. If the infection probabilities are nonzero
on an extensive number of nodes, $\boldsymbol{\rho}$ is delocalized or extended, implying that $\mathcal{I}_N^{(4)}(\boldsymbol{\rho})$ and $\mathcal{I}_N^{(2)}(\boldsymbol{\rho})$
are of order $\mathcal{O}(N^{0})$. Here, we study the behaviour of these quantities strictly in the limit $N \rightarrow \infty$ by numerically
solving Eq. (\ref{hhjj2}) for $\mathcal{P}(\rho)$.

In the endemic phase, the disease infects a finite fraction of individuals and the above parameters converge to
\begin{equation}
  \mathcal{I}^{(4)}(\rho) = \lim_{N \rightarrow \infty} \mathcal{I}_N^{(4)}(\boldsymbol{\rho}) =  \dfrac{\langle \rho^4  \rangle}{ \langle \rho^2   \rangle^2 }
  \label{huhu55}
\end{equation}  
and
\begin{equation}
  \mathcal{I}^{(2)}(\rho) = \lim_{N \rightarrow \infty} \mathcal{I}_N^{(2)}(\boldsymbol{\rho}) =  \dfrac{\langle \rho^2  \rangle}{ \langle \rho   \rangle^2 }.
  \label{huhu56}
\end{equation}  
In figure \ref{phasediagram}, we quantify the fluctuations of the infection probabilities in the endemic phase by
displaying $\mathcal{I}^{(4)}(\rho)$ in a colour scale.
While the IPR can increase by several orders
of magnitude near the phase transition, $\mathcal{I}^{(4)}(\rho)$ remains finite, indicating that the state vector $\boldsymbol{\rho}$ is delocalized in the endemic phase.

Nevertheless, the prevalence $\langle \rho \rangle$ vanishes continuously as we approach
the critical line, suggesting that the disease may become localized 
near the epidemic threshold.
To examine this scenario in more detail, we focus
on the regime $\sigma < \sigma_{*}$, where the epidemic threshold is known
analytically. Moreover, for $0 < c-\lambda^{-1} \ll 1$ ($\sigma < \sigma_{*}$), $\boldsymbol{\rho}$ is governed
by the leading eigenvector $\boldsymbol{R}_1$, responsible
for destabilizing the absorbing phase. The moments of $\boldsymbol{R}_1$ have been analytically computed for directed
complex networks in the limit $N \rightarrow \infty$ \cite{Metz2021}.
\begin{figure}[H]
  \begin{center}
    \includegraphics[scale=0.39]{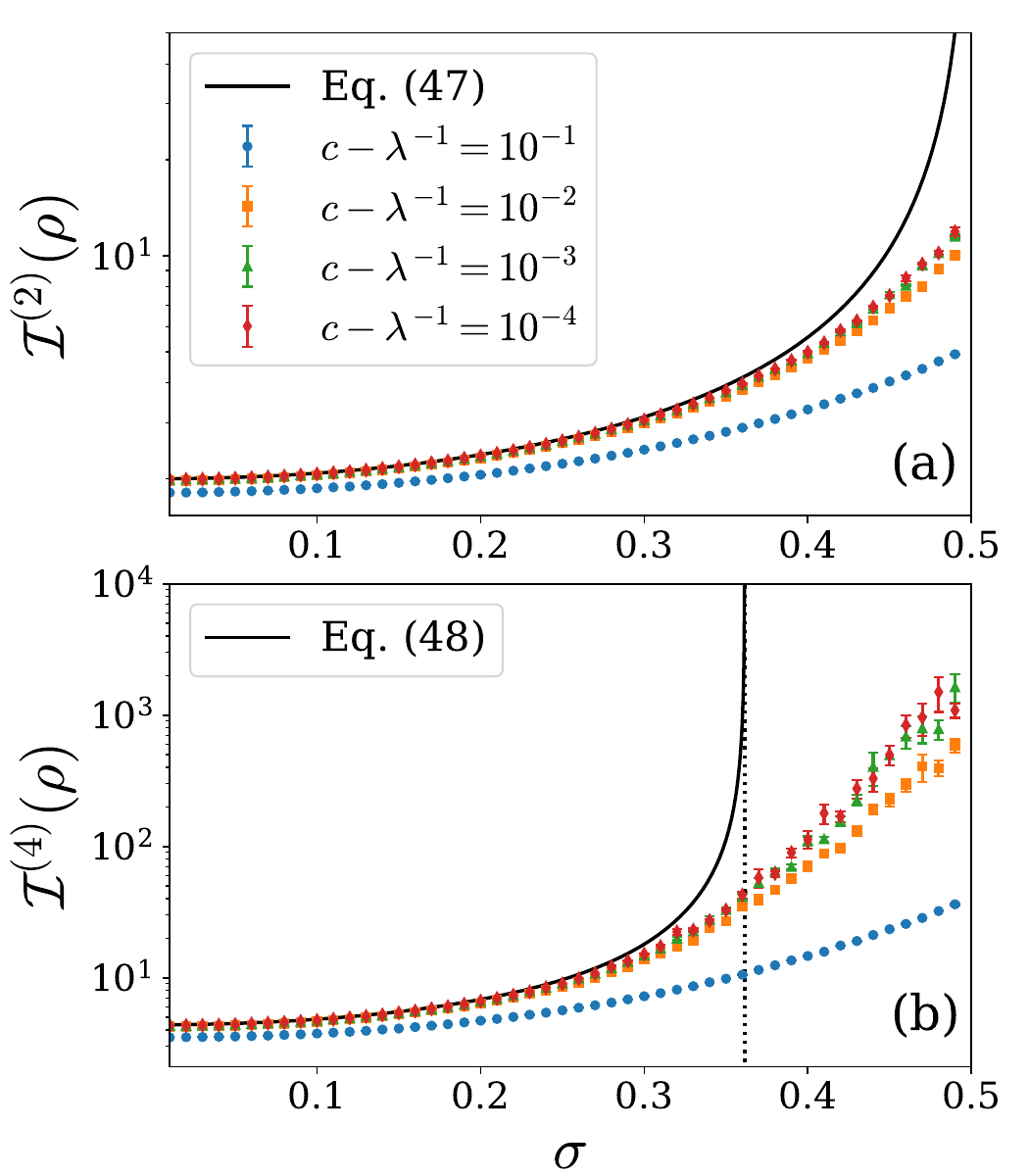} 
    \caption{Dimensionless second moment $\mathcal{I}^{(2)}(\rho)$ [Eq. (\ref{huhu56})] and
      inverse participation ratio $\mathcal{I}^{(4)}(\rho)$ [Eq. (\ref{huhu55})] as
      functions of the standard deviation $\sigma$ of the infection rates for different
      mean degrees $c$ close to the epidemic threshold $c=\lambda^{-1}$. These results are for directed networks with Poisson indegrees and a $\Gamma$-distribution
      of infection rates or coupling strengths with mean $\lambda = 1/2$. Symbols represent numerical results obtained from Eq. (\ref{hhjj2}) using the population
      dynamics algorithm with $M=10^{6}$ (vertical bars indicate the standard deviation of the mean computed over $10$ independent runs). Solid lines are analytic
      predictions derived from the moments of the leading eigenvector [Eqs. (\ref{r1}) and (\ref{r2})].
}
\label{IPR}
\end{center}
\end{figure}
For $\sigma < \sigma_{*}$, the leading eigenvalue is an outlier and the
ratio $\mathcal{I}^{(2)}(R_1)$ of the first two moments of $\boldsymbol{R}_1$ for a Poisson indegree distribution fulfills \cite{Metz2016,Metz2019,Neri2020}
\begin{equation}
  \mathcal{I}^{(2)}(R_1) = \dfrac{c}{c - 1 - \sigma^2/\lambda^2},
  \label{r1}
\end{equation}  
while the IPR  of $\boldsymbol{R}_1$ is given by
\begin{equation}
  \mathcal{I}^{(4)}(R_1) =
  \dfrac{c^3 \left( 3  \mu_3 + 4  c  \mu_ 2 + c^2  \right) - 2  c  \mu_2^2 \left( c^2 + 3 \mu_3  \right) }{\left( c^3 - \mu_4 \right) \left( c^2 - \mu_3 \right) },
  \label{r2}
\end{equation}  
where we defined the dimensionless moments
\begin{equation}
  \mu_n = \lambda^{-n} \int_{0}^{\infty} d x P_{\lambda}(x) x^n
  \label{hhjj}
\end{equation}
of the infection rates or coupling strengths. Depending on the network parameters, Eqs. (\ref{r1}) and (\ref{r2}) diverge due to vanishing
denominators \cite{Metz2021}. In particular, $\mathcal{I}^{(4)}(R_1)$ diverges at $c= \mu_{4}^{\frac{1}{3}}$.
This naturally raises the question of whether $\mathcal{I}^{(2)}(\rho)$ and $\mathcal{I}^{(4)}(\rho)$ exhibit a similar behaviour near $c=\lambda^{-1}$.

In figure \ref{IPR}, we show $\mathcal{I}^{(2)}(\rho)$  and $\mathcal{I}^{(4)}(\rho)$ as functions of $\sigma$
near the epidemic threshold. As a comparison, this figure also displays Eqs. (\ref{r1}) and  (\ref{r2}) at $c = \lambda^{-1}$, demonstrating that
$\mathcal{I}^{(2)}(R_1)$  and $\mathcal{I}^{(4)}(R_1)$ diverge at $\sigma_{*}=\sqrt{\lambda(1-\lambda)}$
and $\sigma_4 \simeq 0.36$, respectively, for the $\Gamma$-distribution of infection rates. The results for $\mathcal{I}^{(2)}(\rho)$  and $\mathcal{I}^{(4)}(\rho)$
are overall
consistent with the analytic predictions obtained from the moments of $\boldsymbol{R}_1$.
The discrepancies observed in figure \ref{IPR} stem from the slow convergence of the population dynamics algorithm near
the thresholds $\sigma_{*}$ and $\sigma_4$, which makes it difficult to
accurately determine the distribution $\mathcal{P}(\rho)$ from the solutions of Eq. (\ref{hhjj2}), as shown in appendix \ref{append}.
In addition, we note that $\mathcal{I}^{(4)}(\rho)$ remains bounded
for $\sigma_4 < \sigma < \sigma_{*}$, a consequence of the finite number of stochastic variables used to discretize $\mathcal{P}(\rho)$ in the population dynamics method.
\begin{figure}
  \begin{center}
    \includegraphics[scale=0.35]{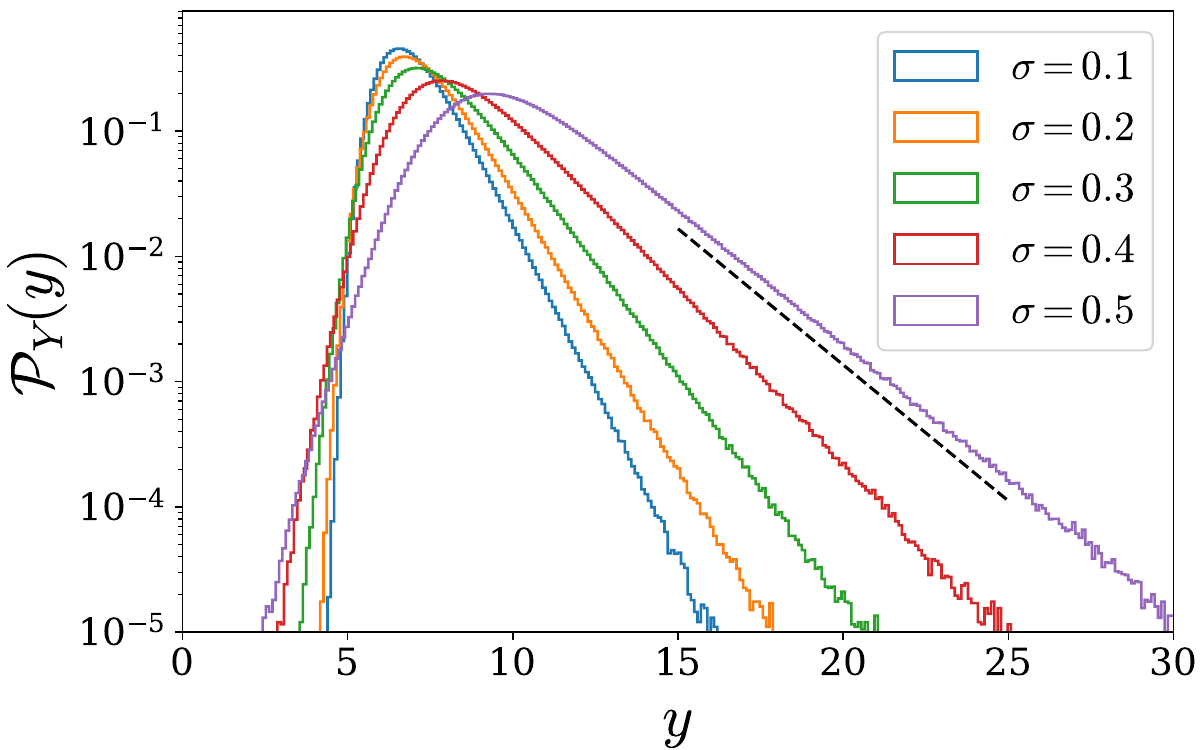} 
    \caption{Probability density $\mathcal{P}_{Y}(y)$ of $Y_i=-\ln \rho_i$ ($i=1,\dots,N$) near the epidemic threshold ($c - \lambda^{-1} = 10^{-4}$). The directed
      network is characterized by a Poisson indegree distribution and a $\Gamma$-distribution
      of infection rates with mean $\lambda = 1/2$ and varying standard deviation $\sigma$. The results are obtained from
      the numerical solutions of Eq. (\ref{hhjj2}) using the population
      dynamics algorithm with $M=10^{6}$ stochastic variables. $\mathcal{P}_{Y}(y)$ exhibits an exponential decay
      for large $y$.
}
\label{density}
\end{center}
\end{figure}

While the moments of the infection probabilities vanish as we approach the critical line, the ratios $\mathcal{I}^{(2)}(\rho)$  and $\mathcal{I}^{(4)}(\rho)$
diverge for sufficiently large $\sigma$, suggesting that $\boldsymbol{\rho}$ may become localized slightly above the epidemic threshold. This
divergence arises from a large fraction of nodes with $\rho_i \simeq 0$ and
a small fraction of nodes with comparatively larger $\rho_i$. To gain further insight into the fluctuations of $\rho_i$, we show in figure \ref{density} the probability
density $\mathcal{P}_{Y}(y)$ of $Y_i = - \ln \rho_i \in [0,\infty)$ near the epidemic threshold. The exponential decay $\mathcal{P}_{Y}(y) \propto e^{- A y}$ for large $y$ implies
that $\mathcal{P}(\rho) \propto \rho^{A-1}$ for $0 < \rho \ll 1$, where the exponent $A$ depends on the network parameters. The power-law
behaviour of $\mathcal{P}(\rho)$ near $\rho=0$ reflects strong fluctuations of $\rho_i$ spanning several orders of magnitude, yet confined to very small
values of $\rho_i$.


\subsection{Stochastic simulations}

In this section, we compare our theoretical results in the limit $N \rightarrow \infty$, obtained within the QMF approximation, with numerical results
from stochastic simulations of the SIS model \cite{KissBook,Cota2017,Wu2019} on directed networks. The simulations are performed using the well-known
Gillespie algorithm \cite{KissBook}.

In the stochastic simulations of the SIS model, the dynamical state of each node $i$ is represented by a binary variable $s_i \in \{ 0,1 \}$ ($i=1,\dots,N$). If node
  $i$ is infected, then $s_i=1$; otherwise, $s_i=0$. The uniform recovery rate $R_i$ of node $i$ is given by $R_i=s_i$, while the infection rate
  of vertex $i$ reads
\begin{equation}
I_i = (1-s_i) \sum_{j \in \partial_{i}^{\rm in}} \lambda_{ij} s_j,
\end{equation}  
where $\partial_{i}^{\rm in}$ denotes the in-neighbourhood of $i$. The coefficient $\lambda_{ij} >0$ is the directed coupling strength
from node $j$ to $i$, representing the rate at which $j$ infects $i$. The variables $\{ \lambda_{ij} \}_{i,j=1,\dots,N}$ are sampled
from a distribution $P_{\lambda}(x)$ which, together with the degree distribution $p_{k,\ell}$, defines the network ensemble (see section \ref{yuyu}).
Thus, infected nodes recover at a unit and constant rate, while susceptible nodes may become infected through contact with their in-neighbourhood.

The time increment between two consecutive updates in the simulation
is sampled from an exponential distribution with rate equal to the total rate of change, $\sum_{i=1}^N (R_i + I_i)$.
To perform a single update in the system, we first select the type of event according to the total recovery and infection rates.
If the selected event is a recovery, an infected node is chosen at random and its state is updated to susceptible. If the event
is an infection, a susceptible node becomes infected with a probability proportional to its infection rate. To select which susceptible
node becomes infected, we employ a rejection sampling method, i.e., we choose a node at random and accept
the choice with probability $I_i/\sum_{j=1}^N I_j$.
The prevalence at a given time is computed as
\begin{equation}
  \langle s \rangle = \frac{1}{N} \sum_{i=1}^N s_i.
  \label{uyuy3}
\end{equation}
The stochastic dynamics of the SIS model on a finite network always converges to the absorbing state for sufficiently long times \cite{Sander2016,Cota2017}, which leads to
computational difficulties in determining the epidemic threshold. To circumvent this issue, we employ the quasistationary
method \cite{Sander2016}, where the prevalence at time $t$ is computed by averaging Eq. (\ref{uyuy3}) over the subset of realizations
that have not reached the absorbing state.

\begin{figure}[H]
  \begin{center}
    \includegraphics[scale=0.34]{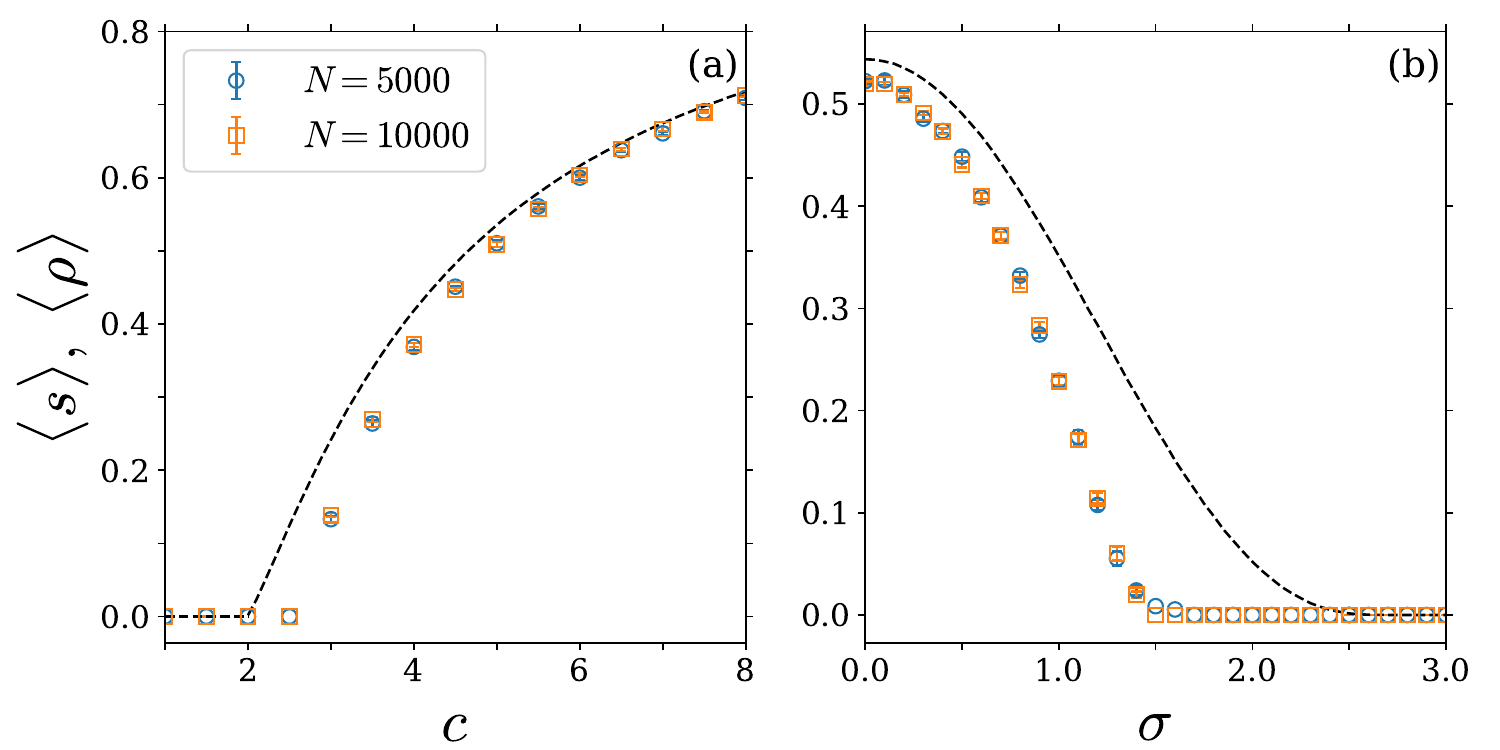} 
    \caption{Comparison between theoretical results in the limit $N \rightarrow \infty$ (dashed lines) and stochastic
      simulation results (symbols) for the stationary prevalence of the SIS model
      on directed networks. The indegrees and outdegrees are independently drawn from a Poisson distribution with mean $c$, while the infection
      rates or coupling strengths $\lambda_{ij}$ follow a $\Gamma$-distribution with mean  $\lambda=1/2$ and standard deviation $\sigma$. The theoretical results
      are obtained by solving Eq. (\ref{hhjj2}) using the population dynamics
      algorithm with $M=10^6$ stochastic variables, while the simulation results are computed by averaging the
      prevalence over $10$ independent realizations after a long simulation time. The simulations are performed using the Gillespie algorithm (see
      the main text). Left panel (a): $\sigma=0.2$. Right panel (b): $c=5$.
}
\label{simula1}
\end{center}
\end{figure}

Figure \ref{simula1} compares theoretical results obtained from the solutions of Eq. (\ref{hhjj2})
with stochastic simulations for different system sizes $N$. Finite size effects are small in the simulations, and the
theoretical predictions for the prevalence, based on the QMF approximation, are in qualitative agreement
with the simulation results. Since the QMF equations neglect pairwise dynamical correlations, the prevalence
computed from Eq. (\ref{hhjj2}) is systematically larger than the simulation results. Accordingly, the theoretical epidemic threshold
is consistently smaller than the one observed in simulations.
Nevertheless, figure \ref{simula1} shows that the agreement between simulations and the QMF approximation
improves as either $c$ increases or $\sigma$ decreases, i.e., in regimes where fluctuations in the contact network structure become
less pronounced. 

\begin{figure}[H]
  \begin{center}
    \includegraphics[scale=0.34]{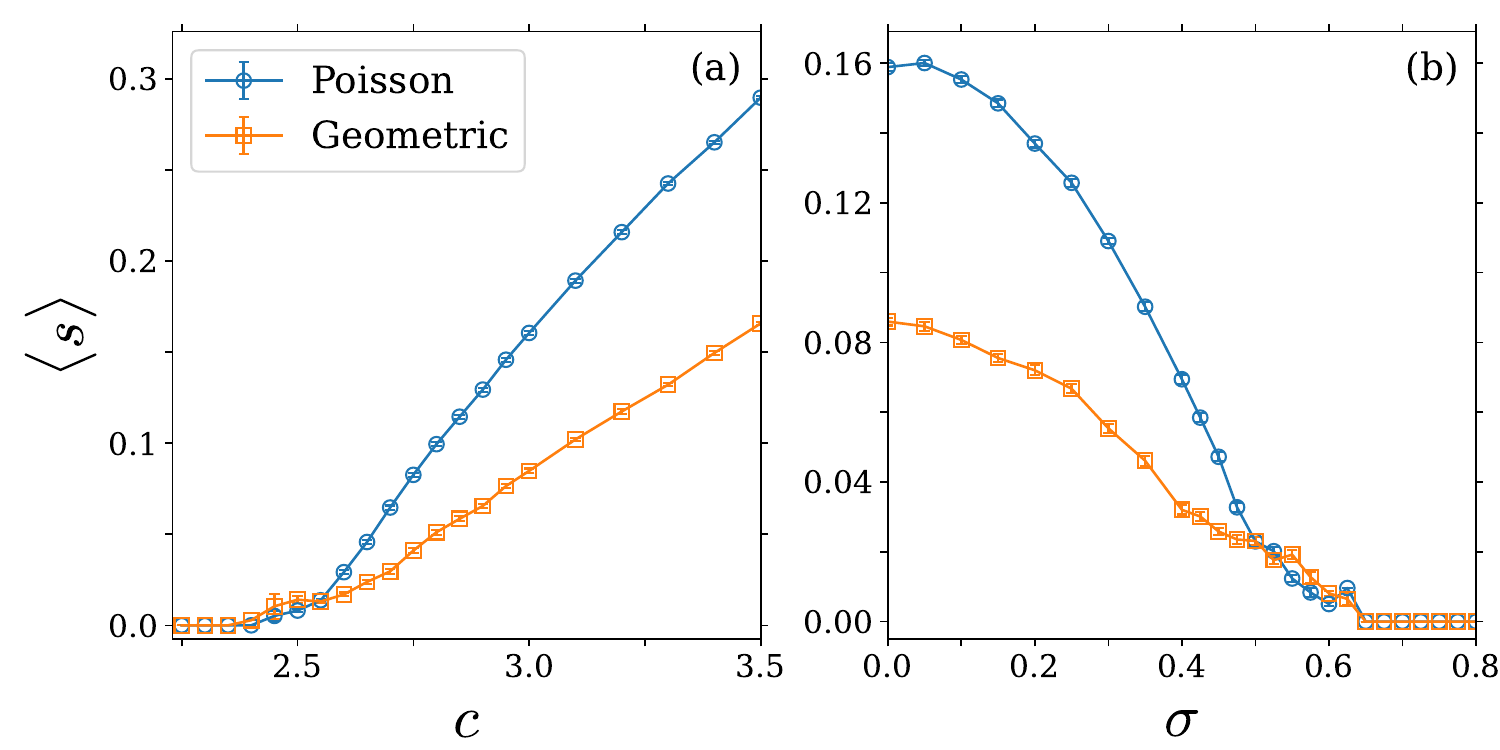} 
    \caption{Stochastic simulation results (symbols) for the prevalence of the SIS model on
        directed networks with two different degree distributions and a $\Gamma$-distribution
      of infection rates with mean $\lambda=1/2$ and standard deviation $\sigma$. The simulations are performed using the Gillespie algorithm (see
      the main text), and the results are obtained by averaging the
      prevalence over $110$ independent realizations of size $N=10^4$ after a long simulation time.
      The solid lines are a guide to the eye. Left panel (a): $\sigma=0$. Right panel (b): $c=3$.
}
\label{simula2}
\end{center}
\end{figure}

In figure \ref{simula2}, we present simulation results for the prevalence across the epidemic threshold
for $N=10^{4}$ and two different degree distributions. These results suggest that the epidemic threshold
is indeed independent of the degree distribution, in qualitative agreement with the predictions derived from Eq. (\ref{hhjj2})
(see figures \ref{transition1} and \ref{transition2}).
Thus, even though our theoretical findings are based on the QMF approximation, Eqs. (\ref{gtrwe}), this simplified description
appears to capture some essential aspects of the phase diagram.


\section{Final remarks}
\label{finalsec}

In this work, we have determined the phase diagram of the SIS model on directed complex networks
within the quenched mean-field approximation. By combining random-matrix results with an analytic
approach for the distribution of stationary infection probabilities, we have computed the epidemic threshold as a function of the mean degree $c$ and
the standard deviation $\sigma$ of the infection rates or coupling strengths defining the contact network.
Our results show that the SIS model exhibits a transition
between the absorbing and endemic phases provided $c \geq \lambda^{-1}$, where $\lambda$ is the average infection rate.

Remarkably, the critical line is independent of the degree distribution but it depends strongly
on the distribution of infection rates.
When $\sigma < \sigma_{*} = \sqrt{\lambda (1-\lambda)}$, both the epidemic threshold and the critical behaviour are governed by the
leading eigenpair of the contact network \cite{Metz2021} and are independent of the infection-rate distribution. In contrast,  for $\sigma > \sigma_{*}$, the fluctuations
of infection rates have a pronounced effect on the critical line. While the SIS model undergoes an absorbing phase transition for
a $\Gamma$-distribution of infection rates as $\sigma$ increases, it remains in the endemic phase for a Pareto distribution.
This striking difference is explained by the large fraction of near-zero infection
rates generated by the $\Gamma$-distribution, which suppresses the endemic state for large $\sigma$.
Together, these results provide a systematic characterization of how network heterogeneity shapes
the phase diagram of the SIS model on directed networks in the quenched mean-field approximation.

We have also examined the emergence of disease localization right above
the epidemic threshold \cite{Goltsev2012}. 
Focusing on the regime where the phase transition is governed by the leading eigenpair of the contact network, we have shown that the
inverse participation ratio (IPR) of the fixed-point infection probabilities diverges near the threshold for sufficiently large $\sigma$, suggesting that the disease becomes localized
on a vanishing fraction of nodes.
These findings are consistent with analytic predictions based on the
IPR of the leading eigenvector \cite{Metz2021}.
We have also computed the full probability density of infection probabilities near the critical line and found
that it exhibits a large fraction of near-zero values, with strong fluctuations spanning several orders of magnitude.

We have compared our theoretical results for $N \rightarrow \infty$ with
stochastic simulations performed using the Gillespie algorithm \cite{KissBook,Cota2017}. The epidemic threshold
computed from the solutions of Eq. (\ref{hhjj2}) consistently lies below the value
obtained from simulations. This quantitative discrepancy is not unexpected, since Eq. (\ref{hhjj2}) is derived
from the quenched mean-field (QMF) approximation, which neglects pairwise dynamical correlations \cite{Mata2013}.
Nevertheless, the simulation results suggest that the exact epidemic threshold
is independent of the degree distribution, in agreement with our theoretical predictions. Overall, this comparison
with stochastic simulations indicates that the QMF approximation captures the main qualitative features of the phase diagram.

We remark that our results are strictly valid in
the limit $N \rightarrow \infty$, as they rely on the numerical
solutions of Eq. (\ref{hhjj2}). Consequently, our approach does not address
how the infection probabilities
or the number of potentially infected nodes in the QMF theory scale with the system size, which could in principle clarify, for instance, why the dimensionless
second moment remains finite while the IPR diverges. A detailed analysis of this issue and its relation with the epidemic survival time
  are interesting directions for future work. Furthermore, it would be interesting to unveil the mechanism
  sustaining  the endemic state in directed graphs and compute the finite-size epidemic threshold by developing, for instance, a directed version
  of the scaling theory in  \cite{Castellano2020}.

The SIS model on synthetic networks provides a theoretical framework to
  qualitatively investigate the effect of the heterogeneous structure of contact networks on epidemic spreading. However, we emphasize that, from
 a practical standpoint, this setting
 is highly idealized and insufficient for making quantitative predictions about real-world spreading phenomena.
From a methodological perspective, we have introduced a general analytic framework to obtain an equation for the
distribution of fixed-point states in the limit $N \rightarrow \infty$. The method is not restricted to the SIS model and can be extended to
study the non-equilibrium fixed-points of other dynamical systems on directed networks, including
the Kuramoto model \cite{Acebron2005} and firing-rate models of neural networks \cite{Sompolinsky1988}. Moreover, since the approach is
based on the cavity method, it can also incorporate other important network features, such as short loops and degree-degree correlations, opening 
promising avenues of future research.

\begin{acknowledgments}  
  V. B. M. acknowledges fellowships from CNPq/Brazil and BIC-UFRGS. F. L. M. acknowledges support from
  CNPq (Grant No 402487/2023-0), FAPERJ (Grant No 204.646/2024), and from ICTP through the Associates Program (2023-2028)
\end{acknowledgments}

\bibliography{biblio}

\begin{thebibliography}{50}
\expandafter\ifx\csname natexlab\endcsname\relax\def\natexlab#1{#1}\fi
\expandafter\ifx\csname bibnamefont\endcsname\relax
  \def\bibnamefont#1{#1}\fi
\expandafter\ifx\csname bibfnamefont\endcsname\relax
  \def\bibfnamefont#1{#1}\fi
\expandafter\ifx\csname citenamefont\endcsname\relax
  \def\citenamefont#1{#1}\fi
\expandafter\ifx\csname url\endcsname\relax
  \def\url#1{\texttt{#1}}\fi
\expandafter\ifx\csname urlprefix\endcsname\relax\def\urlprefix{URL }\fi
\providecommand{\bibinfo}[2]{#2}
\providecommand{\eprint}[2][]{\url{#2}}

\bibitem[{\citenamefont{Kiss et~al.}(2017)\citenamefont{Kiss, Miller, and
  Simon}}]{KissBook}
\bibinfo{author}{\bibfnamefont{I.}~\bibnamefont{Kiss}},
  \bibinfo{author}{\bibfnamefont{J.}~\bibnamefont{Miller}}, \bibnamefont{and}
  \bibinfo{author}{\bibfnamefont{P.}~\bibnamefont{Simon}},
  \emph{\bibinfo{title}{Mathematics of Epidemics on Networks: From Exact to
  Approximate Models}}, Interdisciplinary Applied Mathematics
  (\bibinfo{publisher}{Springer International Publishing},
  \bibinfo{year}{2017}), ISBN \bibinfo{isbn}{9783319508061}.

\bibitem[{\citenamefont{Pastor-Satorras
  et~al.}(2015)\citenamefont{Pastor-Satorras, Castellano, Van~Mieghem, and
  Vespignani}}]{PastorSatorras2015}
\bibinfo{author}{\bibfnamefont{R.}~\bibnamefont{Pastor-Satorras}},
  \bibinfo{author}{\bibfnamefont{C.}~\bibnamefont{Castellano}},
  \bibinfo{author}{\bibfnamefont{P.}~\bibnamefont{Van~Mieghem}},
  \bibnamefont{and}
  \bibinfo{author}{\bibfnamefont{A.}~\bibnamefont{Vespignani}},
  \bibinfo{journal}{Rev. Mod. Phys.} \textbf{\bibinfo{volume}{87}},
  \bibinfo{pages}{925} (\bibinfo{year}{2015}),
  \urlprefix\url{https://link.aps.org/doi/10.1103/RevModPhys.87.925}.

\bibitem[{\citenamefont{Chakrabarti et~al.}(2008)\citenamefont{Chakrabarti,
  Wang, Wang, Leskovec, and Faloutsos}}]{Chakrabarti2008}
\bibinfo{author}{\bibfnamefont{D.}~\bibnamefont{Chakrabarti}},
  \bibinfo{author}{\bibfnamefont{Y.}~\bibnamefont{Wang}},
  \bibinfo{author}{\bibfnamefont{C.}~\bibnamefont{Wang}},
  \bibinfo{author}{\bibfnamefont{J.}~\bibnamefont{Leskovec}}, \bibnamefont{and}
  \bibinfo{author}{\bibfnamefont{C.}~\bibnamefont{Faloutsos}},
  \textbf{\bibinfo{volume}{10}} (\bibinfo{year}{2008}), ISSN
  \bibinfo{issn}{1094-9224},
  \urlprefix\url{https://doi.org/10.1145/1284680.1284681}.

\bibitem[{\citenamefont{Chatterjee and Durrett}(2009)}]{Chatterjee2009}
\bibinfo{author}{\bibfnamefont{S.}~\bibnamefont{Chatterjee}} \bibnamefont{and}
  \bibinfo{author}{\bibfnamefont{R.}~\bibnamefont{Durrett}},
  \bibinfo{journal}{The Annals of Probability} \textbf{\bibinfo{volume}{37}},
  \bibinfo{pages}{2332} (\bibinfo{year}{2009}), ISSN \bibinfo{issn}{00911798},
  \urlprefix\url{http://www.jstor.org/stable/27795079}.

\bibitem[{\citenamefont{Castellano and Pastor-Satorras}(2010)}]{Castellano2010}
\bibinfo{author}{\bibfnamefont{C.}~\bibnamefont{Castellano}} \bibnamefont{and}
  \bibinfo{author}{\bibfnamefont{R.}~\bibnamefont{Pastor-Satorras}},
  \bibinfo{journal}{Phys. Rev. Lett.} \textbf{\bibinfo{volume}{105}},
  \bibinfo{pages}{218701} (\bibinfo{year}{2010}),
  \urlprefix\url{https://link.aps.org/doi/10.1103/PhysRevLett.105.218701}.

\bibitem[{\citenamefont{Ferreira et~al.}(2012)\citenamefont{Ferreira,
  Castellano, and Pastor-Satorras}}]{Ferreira2012}
\bibinfo{author}{\bibfnamefont{S.~C.} \bibnamefont{Ferreira}},
  \bibinfo{author}{\bibfnamefont{C.}~\bibnamefont{Castellano}},
  \bibnamefont{and}
  \bibinfo{author}{\bibfnamefont{R.}~\bibnamefont{Pastor-Satorras}},
  \bibinfo{journal}{Phys. Rev. E} \textbf{\bibinfo{volume}{86}},
  \bibinfo{pages}{041125} (\bibinfo{year}{2012}),
  \urlprefix\url{https://link.aps.org/doi/10.1103/PhysRevE.86.041125}.

\bibitem[{\citenamefont{Goltsev et~al.}(2012)\citenamefont{Goltsev,
  Dorogovtsev, Oliveira, and Mendes}}]{Goltsev2012}
\bibinfo{author}{\bibfnamefont{A.~V.} \bibnamefont{Goltsev}},
  \bibinfo{author}{\bibfnamefont{S.~N.} \bibnamefont{Dorogovtsev}},
  \bibinfo{author}{\bibfnamefont{J.~G.} \bibnamefont{Oliveira}},
  \bibnamefont{and} \bibinfo{author}{\bibfnamefont{J.~F.~F.}
  \bibnamefont{Mendes}}, \bibinfo{journal}{Phys. Rev. Lett.}
  \textbf{\bibinfo{volume}{109}}, \bibinfo{pages}{128702}
  (\bibinfo{year}{2012}),
  \urlprefix\url{https://link.aps.org/doi/10.1103/PhysRevLett.109.128702}.

\bibitem[{\citenamefont{Li et~al.}(2013)\citenamefont{Li, Wang, and
  Van~Mieghem}}]{Li2013}
\bibinfo{author}{\bibfnamefont{C.}~\bibnamefont{Li}},
  \bibinfo{author}{\bibfnamefont{H.}~\bibnamefont{Wang}}, \bibnamefont{and}
  \bibinfo{author}{\bibfnamefont{P.}~\bibnamefont{Van~Mieghem}},
  \bibinfo{journal}{Phys. Rev. E} \textbf{\bibinfo{volume}{88}},
  \bibinfo{pages}{062802} (\bibinfo{year}{2013}),
  \urlprefix\url{https://link.aps.org/doi/10.1103/PhysRevE.88.062802}.

\bibitem[{\citenamefont{Van~Mieghem and van~de
  Bovenkamp}(2013)}]{VanMieghem2013}
\bibinfo{author}{\bibfnamefont{P.}~\bibnamefont{Van~Mieghem}} \bibnamefont{and}
  \bibinfo{author}{\bibfnamefont{R.}~\bibnamefont{van~de Bovenkamp}},
  \bibinfo{journal}{Phys. Rev. Lett.} \textbf{\bibinfo{volume}{110}},
  \bibinfo{pages}{108701} (\bibinfo{year}{2013}),
  \urlprefix\url{https://link.aps.org/doi/10.1103/PhysRevLett.110.108701}.

\bibitem[{\citenamefont{Kwon and Kim}(2013)}]{Kwon2013}
\bibinfo{author}{\bibfnamefont{S.}~\bibnamefont{Kwon}} \bibnamefont{and}
  \bibinfo{author}{\bibfnamefont{Y.}~\bibnamefont{Kim}},
  \bibinfo{journal}{Phys. Rev. E} \textbf{\bibinfo{volume}{87}},
  \bibinfo{pages}{012813} (\bibinfo{year}{2013}),
  \urlprefix\url{https://link.aps.org/doi/10.1103/PhysRevE.87.012813}.

\bibitem[{\citenamefont{Mata and Ferreira}(2013)}]{Mata2013}
\bibinfo{author}{\bibfnamefont{A.~S.} \bibnamefont{Mata}} \bibnamefont{and}
  \bibinfo{author}{\bibfnamefont{S.~C.} \bibnamefont{Ferreira}},
  \bibinfo{journal}{Europhysics Letters} \textbf{\bibinfo{volume}{103}},
  \bibinfo{pages}{48003} (\bibinfo{year}{2013}),
  \urlprefix\url{https://dx.doi.org/10.1209/0295-5075/103/48003}.

\bibitem[{\citenamefont{Newman}(2010)}]{Newman2010}
\bibinfo{author}{\bibfnamefont{M.}~\bibnamefont{Newman}},
  \emph{\bibinfo{title}{Networks: An Introduction}} (\bibinfo{publisher}{OUP
  Oxford}, \bibinfo{year}{2010}), ISBN \bibinfo{isbn}{9780199206650}.

\bibitem[{\citenamefont{Krivelevich and Sudakov}(2003)}]{Krivelevich2003}
\bibinfo{author}{\bibfnamefont{M.}~\bibnamefont{Krivelevich}} \bibnamefont{and}
  \bibinfo{author}{\bibfnamefont{B.}~\bibnamefont{Sudakov}},
  \bibinfo{journal}{Comb. Probab. Comput.} \textbf{\bibinfo{volume}{12}},
  \bibinfo{pages}{61} (\bibinfo{year}{2003}).

\bibitem[{\citenamefont{Chung et~al.}(2003)\citenamefont{Chung, Lu, and
  Vu}}]{Chung2003}
\bibinfo{author}{\bibfnamefont{F.}~\bibnamefont{Chung}},
  \bibinfo{author}{\bibfnamefont{L.}~\bibnamefont{Lu}}, \bibnamefont{and}
  \bibinfo{author}{\bibfnamefont{V.}~\bibnamefont{Vu}},
  \bibinfo{journal}{Proceedings of the National Academy of Sciences}
  \textbf{\bibinfo{volume}{100}}, \bibinfo{pages}{6313} (\bibinfo{year}{2003}),
  \eprint{https://www.pnas.org/doi/pdf/10.1073/pnas.0937490100},
  \urlprefix\url{https://www.pnas.org/doi/abs/10.1073/pnas.0937490100}.

\bibitem[{\citenamefont{Pastor-Satorras and
  Castellano}(2018)}]{PastorSatorras2018}
\bibinfo{author}{\bibfnamefont{R.}~\bibnamefont{Pastor-Satorras}}
  \bibnamefont{and}
  \bibinfo{author}{\bibfnamefont{C.}~\bibnamefont{Castellano}},
  \bibinfo{journal}{J. Stat. Phys.} \textbf{\bibinfo{volume}{173}},
  \bibinfo{pages}{1110} (\bibinfo{year}{2018}).

\bibitem[{\citenamefont{Abou-Chacra et~al.}(1973)\citenamefont{Abou-Chacra,
  Thouless, and Anderson}}]{AbouChacra1973}
\bibinfo{author}{\bibfnamefont{R.}~\bibnamefont{Abou-Chacra}},
  \bibinfo{author}{\bibfnamefont{D.~J.} \bibnamefont{Thouless}},
  \bibnamefont{and} \bibinfo{author}{\bibfnamefont{P.~W.}
  \bibnamefont{Anderson}}, \bibinfo{journal}{Journal of Physics C: Solid State
  Physics} \textbf{\bibinfo{volume}{6}}, \bibinfo{pages}{1734}
  (\bibinfo{year}{1973}),
  \urlprefix\url{https://dx.doi.org/10.1088/0022-3719/6/10/009}.

\bibitem[{\citenamefont{Mirlin}(2000)}]{Mirlin2000}
\bibinfo{author}{\bibfnamefont{A.~D.} \bibnamefont{Mirlin}},
  \bibinfo{journal}{Physics Reports} \textbf{\bibinfo{volume}{326}},
  \bibinfo{pages}{259} (\bibinfo{year}{2000}), ISSN \bibinfo{issn}{0370-1573},
  \urlprefix\url{https://www.sciencedirect.com/science/article/pii/S0370157399000915}.

\bibitem[{\citenamefont{Pastor-Satorras and
  Castellano}(2016)}]{PastorSatorras2016}
\bibinfo{author}{\bibfnamefont{R.}~\bibnamefont{Pastor-Satorras}}
  \bibnamefont{and}
  \bibinfo{author}{\bibfnamefont{C.}~\bibnamefont{Castellano}},
  \bibinfo{journal}{Sci. Rep.} \textbf{\bibinfo{volume}{6}},
  \bibinfo{pages}{18847} (\bibinfo{year}{2016}).

\bibitem[{\citenamefont{Liu and Van~Mieghem}(2019)}]{Liu2019}
\bibinfo{author}{\bibfnamefont{Q.}~\bibnamefont{Liu}} \bibnamefont{and}
  \bibinfo{author}{\bibfnamefont{P.}~\bibnamefont{Van~Mieghem}},
  \bibinfo{journal}{IEEE Transactions on Network Science and Engineering}
  \textbf{\bibinfo{volume}{6}}, \bibinfo{pages}{983} (\bibinfo{year}{2019}).

\bibitem[{\citenamefont{Newman et~al.}(2002)\citenamefont{Newman, Forrest, and
  Balthrop}}]{Newman2002}
\bibinfo{author}{\bibfnamefont{M.~E.~J.} \bibnamefont{Newman}},
  \bibinfo{author}{\bibfnamefont{S.}~\bibnamefont{Forrest}}, \bibnamefont{and}
  \bibinfo{author}{\bibfnamefont{J.}~\bibnamefont{Balthrop}},
  \bibinfo{journal}{Phys. Rev. E} \textbf{\bibinfo{volume}{66}},
  \bibinfo{pages}{035101} (\bibinfo{year}{2002}),
  \urlprefix\url{https://link.aps.org/doi/10.1103/PhysRevE.66.035101}.

\bibitem[{\citenamefont{Ebel et~al.}(2002)\citenamefont{Ebel, Mielsch, and
  Bornholdt}}]{Ebel2002}
\bibinfo{author}{\bibfnamefont{H.}~\bibnamefont{Ebel}},
  \bibinfo{author}{\bibfnamefont{L.-I.} \bibnamefont{Mielsch}},
  \bibnamefont{and}
  \bibinfo{author}{\bibfnamefont{S.}~\bibnamefont{Bornholdt}},
  \bibinfo{journal}{Phys. Rev. E} \textbf{\bibinfo{volume}{66}},
  \bibinfo{pages}{035103} (\bibinfo{year}{2002}),
  \urlprefix\url{https://link.aps.org/doi/10.1103/PhysRevE.66.035103}.

\bibitem[{\citenamefont{Krueger et~al.}(2024)\citenamefont{Krueger, Mitra,
  Ozanski, and Pramanik}}]{Krueger2024}
\bibinfo{author}{\bibfnamefont{T.}~\bibnamefont{Krueger}},
  \bibinfo{author}{\bibfnamefont{B.}~\bibnamefont{Mitra}},
  \bibinfo{author}{\bibfnamefont{T.}~\bibnamefont{Ozanski}}, \bibnamefont{and}
  \bibinfo{author}{\bibfnamefont{S.}~\bibnamefont{Pramanik}},
  \bibinfo{journal}{IEEE Transactions on Network Science and Engineering}
  \textbf{\bibinfo{volume}{11}}, \bibinfo{pages}{2742} (\bibinfo{year}{2024}).

\bibitem[{\citenamefont{Meyers et~al.}(2006)\citenamefont{Meyers, Newman, and
  Pourbohloul}}]{Meyers2006}
\bibinfo{author}{\bibfnamefont{L.~A.} \bibnamefont{Meyers}},
  \bibinfo{author}{\bibfnamefont{M.}~\bibnamefont{Newman}}, \bibnamefont{and}
  \bibinfo{author}{\bibfnamefont{B.}~\bibnamefont{Pourbohloul}},
  \bibinfo{journal}{Journal of Theoretical Biology}
  \textbf{\bibinfo{volume}{240}}, \bibinfo{pages}{400} (\bibinfo{year}{2006}),
  ISSN \bibinfo{issn}{0022-5193},
  \urlprefix\url{https://www.sciencedirect.com/science/article/pii/S0022519305004418}.

\bibitem[{\citenamefont{Metz and P\'erez~Castillo}(2016)}]{Metz2016}
\bibinfo{author}{\bibfnamefont{F.~L.} \bibnamefont{Metz}} \bibnamefont{and}
  \bibinfo{author}{\bibfnamefont{I.}~\bibnamefont{P\'erez~Castillo}},
  \bibinfo{journal}{Phys. Rev. Lett.} \textbf{\bibinfo{volume}{117}},
  \bibinfo{pages}{104101} (\bibinfo{year}{2016}),
  \urlprefix\url{https://link.aps.org/doi/10.1103/PhysRevLett.117.104101}.

\bibitem[{\citenamefont{Lucas~Metz et~al.}(2019)\citenamefont{Lucas~Metz, Neri,
  and Rogers}}]{Metz2019}
\bibinfo{author}{\bibfnamefont{F.}~\bibnamefont{Lucas~Metz}},
  \bibinfo{author}{\bibfnamefont{I.}~\bibnamefont{Neri}}, \bibnamefont{and}
  \bibinfo{author}{\bibfnamefont{T.}~\bibnamefont{Rogers}},
  \bibinfo{journal}{Journal of Physics A: Mathematical and Theoretical}
  \textbf{\bibinfo{volume}{52}}, \bibinfo{pages}{434003}
  (\bibinfo{year}{2019}),
  \urlprefix\url{https://dx.doi.org/10.1088/1751-8121/ab1ce0}.

\bibitem[{\citenamefont{Metz and Neri}(2021)}]{Metz2021}
\bibinfo{author}{\bibfnamefont{F.~L.} \bibnamefont{Metz}} \bibnamefont{and}
  \bibinfo{author}{\bibfnamefont{I.}~\bibnamefont{Neri}},
  \bibinfo{journal}{Phys. Rev. Lett.} \textbf{\bibinfo{volume}{126}},
  \bibinfo{pages}{040604} (\bibinfo{year}{2021}),
  \urlprefix\url{https://link.aps.org/doi/10.1103/PhysRevLett.126.040604}.

\bibitem[{\citenamefont{M\'ezard and Parisi}(2001)}]{Mezard2001}
\bibinfo{author}{\bibfnamefont{M.}~\bibnamefont{M\'ezard}} \bibnamefont{and}
  \bibinfo{author}{\bibfnamefont{G.}~\bibnamefont{Parisi}},
  \bibinfo{journal}{Eur. Phys. J. B} \textbf{\bibinfo{volume}{20}},
  \bibinfo{pages}{217} (\bibinfo{year}{2001}).

\bibitem[{\citenamefont{Metz et~al.}(2010)\citenamefont{Metz, Neri, and
  Boll\'e}}]{Metz2010}
\bibinfo{author}{\bibfnamefont{F.~L.} \bibnamefont{Metz}},
  \bibinfo{author}{\bibfnamefont{I.}~\bibnamefont{Neri}}, \bibnamefont{and}
  \bibinfo{author}{\bibfnamefont{D.}~\bibnamefont{Boll\'e}},
  \bibinfo{journal}{Phys. Rev. E} \textbf{\bibinfo{volume}{82}},
  \bibinfo{pages}{031135} (\bibinfo{year}{2010}),
  \urlprefix\url{https://link.aps.org/doi/10.1103/PhysRevE.82.031135}.

\bibitem[{\citenamefont{Newman et~al.}(2001)\citenamefont{Newman, Strogatz, and
  Watts}}]{Newman2001}
\bibinfo{author}{\bibfnamefont{M.~E.~J.} \bibnamefont{Newman}},
  \bibinfo{author}{\bibfnamefont{S.~H.} \bibnamefont{Strogatz}},
  \bibnamefont{and} \bibinfo{author}{\bibfnamefont{D.~J.} \bibnamefont{Watts}},
  \bibinfo{journal}{Phys. Rev. E} \textbf{\bibinfo{volume}{64}},
  \bibinfo{pages}{026118} (\bibinfo{year}{2001}),
  \urlprefix\url{https://link.aps.org/doi/10.1103/PhysRevE.64.026118}.

\bibitem[{\citenamefont{Fosdick et~al.}(2018)\citenamefont{Fosdick, Larremore,
  Nishimura, and Ugander}}]{Fosdick2018}
\bibinfo{author}{\bibfnamefont{B.~K.} \bibnamefont{Fosdick}},
  \bibinfo{author}{\bibfnamefont{D.~B.} \bibnamefont{Larremore}},
  \bibinfo{author}{\bibfnamefont{J.}~\bibnamefont{Nishimura}},
  \bibnamefont{and} \bibinfo{author}{\bibfnamefont{J.}~\bibnamefont{Ugander}},
  \bibinfo{journal}{SIAM Review} \textbf{\bibinfo{volume}{60}},
  \bibinfo{pages}{315} (\bibinfo{year}{2018}),
  \eprint{https://doi.org/10.1137/16M1087175},
  \urlprefix\url{https://doi.org/10.1137/16M1087175}.

\bibitem[{\citenamefont{Kryven}(2016)}]{Kryven2016}
\bibinfo{author}{\bibfnamefont{I.}~\bibnamefont{Kryven}},
  \bibinfo{journal}{Phys. Rev. E} \textbf{\bibinfo{volume}{94}},
  \bibinfo{pages}{012315} (\bibinfo{year}{2016}),
  \urlprefix\url{https://link.aps.org/doi/10.1103/PhysRevE.94.012315}.

\bibitem[{\citenamefont{Neri and Metz}(2020)}]{Neri2020}
\bibinfo{author}{\bibfnamefont{I.}~\bibnamefont{Neri}} \bibnamefont{and}
  \bibinfo{author}{\bibfnamefont{F.~L.} \bibnamefont{Metz}},
  \bibinfo{journal}{Phys. Rev. Res.} \textbf{\bibinfo{volume}{2}},
  \bibinfo{pages}{033313} (\bibinfo{year}{2020}),
  \urlprefix\url{https://link.aps.org/doi/10.1103/PhysRevResearch.2.033313}.

\bibitem[{\citenamefont{Metz}(2025)}]{Metz2025}
\bibinfo{author}{\bibfnamefont{F.~L.} \bibnamefont{Metz}},
  \bibinfo{journal}{Phys. Rev. Lett.} \textbf{\bibinfo{volume}{134}},
  \bibinfo{pages}{037401} (\bibinfo{year}{2025}),
  \urlprefix\url{https://link.aps.org/doi/10.1103/PhysRevLett.134.037401}.

\bibitem[{\citenamefont{M{\'e}zard and Montanari}(2009)}]{MezardBook}
\bibinfo{author}{\bibfnamefont{M.}~\bibnamefont{M{\'e}zard}} \bibnamefont{and}
  \bibinfo{author}{\bibfnamefont{A.}~\bibnamefont{Montanari}},
  \emph{\bibinfo{title}{Information, Physics, and Computation}}, Oxford
  Graduate Texts (\bibinfo{publisher}{OUP Oxford}, \bibinfo{year}{2009}), ISBN
  \bibinfo{isbn}{9780198570837}.

\bibitem[{\citenamefont{Lokhov et~al.}(2014)\citenamefont{Lokhov, M\'ezard,
  Ohta, and Zdeborov\'a}}]{Lokhov2014}
\bibinfo{author}{\bibfnamefont{A.~Y.} \bibnamefont{Lokhov}},
  \bibinfo{author}{\bibfnamefont{M.}~\bibnamefont{M\'ezard}},
  \bibinfo{author}{\bibfnamefont{H.}~\bibnamefont{Ohta}}, \bibnamefont{and}
  \bibinfo{author}{\bibfnamefont{L.}~\bibnamefont{Zdeborov\'a}},
  \bibinfo{journal}{Phys. Rev. E} \textbf{\bibinfo{volume}{90}},
  \bibinfo{pages}{012801} (\bibinfo{year}{2014}),
  \urlprefix\url{https://link.aps.org/doi/10.1103/PhysRevE.90.012801}.

\bibitem[{\citenamefont{Lokhov et~al.}(2015)\citenamefont{Lokhov, M\'ezard, and
  Zdeborov\'a}}]{Lokhov2015}
\bibinfo{author}{\bibfnamefont{A.~Y.} \bibnamefont{Lokhov}},
  \bibinfo{author}{\bibfnamefont{M.}~\bibnamefont{M\'ezard}}, \bibnamefont{and}
  \bibinfo{author}{\bibfnamefont{L.}~\bibnamefont{Zdeborov\'a}},
  \bibinfo{journal}{Phys. Rev. E} \textbf{\bibinfo{volume}{91}},
  \bibinfo{pages}{012811} (\bibinfo{year}{2015}),
  \urlprefix\url{https://link.aps.org/doi/10.1103/PhysRevE.91.012811}.

\bibitem[{\citenamefont{Shrestha et~al.}(2015)\citenamefont{Shrestha, Scarpino,
  and Moore}}]{Shrestha2015}
\bibinfo{author}{\bibfnamefont{M.}~\bibnamefont{Shrestha}},
  \bibinfo{author}{\bibfnamefont{S.~V.} \bibnamefont{Scarpino}},
  \bibnamefont{and} \bibinfo{author}{\bibfnamefont{C.}~\bibnamefont{Moore}},
  \bibinfo{journal}{Phys. Rev. E} \textbf{\bibinfo{volume}{92}},
  \bibinfo{pages}{022821} (\bibinfo{year}{2015}).

\bibitem[{\citenamefont{Ortega et~al.}(2022)\citenamefont{Ortega, Machado, and
  Lage-Castellanos}}]{Ortega2022}
\bibinfo{author}{\bibfnamefont{E.}~\bibnamefont{Ortega}},
  \bibinfo{author}{\bibfnamefont{D.}~\bibnamefont{Machado}}, \bibnamefont{and}
  \bibinfo{author}{\bibfnamefont{A.}~\bibnamefont{Lage-Castellanos}},
  \bibinfo{journal}{Phys. Rev. E} \textbf{\bibinfo{volume}{105}},
  \bibinfo{pages}{024308} (\bibinfo{year}{2022}),
  \urlprefix\url{https://link.aps.org/doi/10.1103/PhysRevE.105.024308}.

\bibitem[{\citenamefont{Altarelli
  et~al.}(2014{\natexlab{a}})\citenamefont{Altarelli, Braunstein, Dall'Asta,
  Lage-Castellanos, and Zecchina}}]{Altarelli2014}
\bibinfo{author}{\bibfnamefont{F.}~\bibnamefont{Altarelli}},
  \bibinfo{author}{\bibfnamefont{A.}~\bibnamefont{Braunstein}},
  \bibinfo{author}{\bibfnamefont{L.}~\bibnamefont{Dall'Asta}},
  \bibinfo{author}{\bibfnamefont{A.}~\bibnamefont{Lage-Castellanos}},
  \bibnamefont{and} \bibinfo{author}{\bibfnamefont{R.}~\bibnamefont{Zecchina}},
  \bibinfo{journal}{Phys. Rev. Lett.} \textbf{\bibinfo{volume}{112}},
  \bibinfo{pages}{118701} (\bibinfo{year}{2014}{\natexlab{a}}),
  \urlprefix\url{https://link.aps.org/doi/10.1103/PhysRevLett.112.118701}.

\bibitem[{\citenamefont{Altarelli
  et~al.}(2014{\natexlab{b}})\citenamefont{Altarelli, Braunstein, Dall'Asta,
  Wakeling, and Zecchina}}]{Altarelli2014A}
\bibinfo{author}{\bibfnamefont{F.}~\bibnamefont{Altarelli}},
  \bibinfo{author}{\bibfnamefont{A.}~\bibnamefont{Braunstein}},
  \bibinfo{author}{\bibfnamefont{L.}~\bibnamefont{Dall'Asta}},
  \bibinfo{author}{\bibfnamefont{J.~R.} \bibnamefont{Wakeling}},
  \bibnamefont{and} \bibinfo{author}{\bibfnamefont{R.}~\bibnamefont{Zecchina}},
  \bibinfo{journal}{Phys. Rev. X} \textbf{\bibinfo{volume}{4}},
  \bibinfo{pages}{021024} (\bibinfo{year}{2014}{\natexlab{b}}),
  \urlprefix\url{https://link.aps.org/doi/10.1103/PhysRevX.4.021024}.

\bibitem[{\citenamefont{Kurchan}(1991)}]{Kurchan1991}
\bibinfo{author}{\bibfnamefont{J.}~\bibnamefont{Kurchan}},
  \bibinfo{journal}{Journal of Physics A: Mathematical and General}
  \textbf{\bibinfo{volume}{24}}, \bibinfo{pages}{4969} (\bibinfo{year}{1991}),
  \urlprefix\url{https://dx.doi.org/10.1088/0305-4470/24/21/011}.

\bibitem[{\citenamefont{Neri and Metz}(2016)}]{Neri2016}
\bibinfo{author}{\bibfnamefont{I.}~\bibnamefont{Neri}} \bibnamefont{and}
  \bibinfo{author}{\bibfnamefont{F.~L.} \bibnamefont{Metz}},
  \bibinfo{journal}{Phys. Rev. Lett.} \textbf{\bibinfo{volume}{117}},
  \bibinfo{pages}{224101} (\bibinfo{year}{2016}),
  \urlprefix\url{https://link.aps.org/doi/10.1103/PhysRevLett.117.224101}.

\bibitem[{\citenamefont{Horn and Johnson}(2012)}]{Horn2012}
\bibinfo{author}{\bibfnamefont{R.}~\bibnamefont{Horn}} \bibnamefont{and}
  \bibinfo{author}{\bibfnamefont{C.}~\bibnamefont{Johnson}},
  \emph{\bibinfo{title}{Matrix Analysis}} (\bibinfo{publisher}{Cambridge
  University Press}, \bibinfo{year}{2012}), ISBN \bibinfo{isbn}{9781139788885}.

\bibitem[{\citenamefont{\'Odor}(2004)}]{Odor2004}
\bibinfo{author}{\bibfnamefont{G.}~\bibnamefont{\'Odor}},
  \bibinfo{journal}{Rev. Mod. Phys.} \textbf{\bibinfo{volume}{76}},
  \bibinfo{pages}{663} (\bibinfo{year}{2004}),
  \urlprefix\url{https://link.aps.org/doi/10.1103/RevModPhys.76.663}.

\bibitem[{\citenamefont{Cota and Ferreira}(2017)}]{Cota2017}
\bibinfo{author}{\bibfnamefont{W.}~\bibnamefont{Cota}} \bibnamefont{and}
  \bibinfo{author}{\bibfnamefont{S.~C.} \bibnamefont{Ferreira}},
  \bibinfo{journal}{Computer Physics Communications}
  \textbf{\bibinfo{volume}{219}}, \bibinfo{pages}{303} (\bibinfo{year}{2017}),
  ISSN \bibinfo{issn}{0010-4655},
  \urlprefix\url{https://www.sciencedirect.com/science/article/pii/S0010465517301893}.

\bibitem[{\citenamefont{Wu et~al.}(2019)\citenamefont{Wu, Zhou, and
  Hadzibeganovic}}]{Wu2019}
\bibinfo{author}{\bibfnamefont{Q.}~\bibnamefont{Wu}},
  \bibinfo{author}{\bibfnamefont{R.}~\bibnamefont{Zhou}}, \bibnamefont{and}
  \bibinfo{author}{\bibfnamefont{T.}~\bibnamefont{Hadzibeganovic}},
  \bibinfo{journal}{Physica A: Statistical Mechanics and its Applications}
  \textbf{\bibinfo{volume}{518}}, \bibinfo{pages}{71} (\bibinfo{year}{2019}),
  ISSN \bibinfo{issn}{0378-4371},
  \urlprefix\url{https://www.sciencedirect.com/science/article/pii/S0378437118314894}.

\bibitem[{\citenamefont{Sander et~al.}(2016)\citenamefont{Sander, Costa, and
  Ferreira}}]{Sander2016}
\bibinfo{author}{\bibfnamefont{R.~S.} \bibnamefont{Sander}},
  \bibinfo{author}{\bibfnamefont{G.~S.} \bibnamefont{Costa}}, \bibnamefont{and}
  \bibinfo{author}{\bibfnamefont{S.~C.} \bibnamefont{Ferreira}},
  \bibinfo{journal}{Phys. Rev. E} \textbf{\bibinfo{volume}{94}},
  \bibinfo{pages}{042308} (\bibinfo{year}{2016}),
  \urlprefix\url{https://link.aps.org/doi/10.1103/PhysRevE.94.042308}.

\bibitem[{\citenamefont{Castellano and Pastor-Satorras}(2020)}]{Castellano2020}
\bibinfo{author}{\bibfnamefont{C.}~\bibnamefont{Castellano}} \bibnamefont{and}
  \bibinfo{author}{\bibfnamefont{R.}~\bibnamefont{Pastor-Satorras}},
  \bibinfo{journal}{Phys. Rev. X} \textbf{\bibinfo{volume}{10}},
  \bibinfo{pages}{011070} (\bibinfo{year}{2020}),
  \urlprefix\url{https://link.aps.org/doi/10.1103/PhysRevX.10.011070}.

\bibitem[{\citenamefont{Acebr\'on et~al.}(2005)\citenamefont{Acebr\'on,
  Bonilla, P\'erez~Vicente, Ritort, and Spigler}}]{Acebron2005}
\bibinfo{author}{\bibfnamefont{J.~A.} \bibnamefont{Acebr\'on}},
  \bibinfo{author}{\bibfnamefont{L.~L.} \bibnamefont{Bonilla}},
  \bibinfo{author}{\bibfnamefont{C.~J.} \bibnamefont{P\'erez~Vicente}},
  \bibinfo{author}{\bibfnamefont{F.}~\bibnamefont{Ritort}}, \bibnamefont{and}
  \bibinfo{author}{\bibfnamefont{R.}~\bibnamefont{Spigler}},
  \bibinfo{journal}{Rev. Mod. Phys.} \textbf{\bibinfo{volume}{77}},
  \bibinfo{pages}{137} (\bibinfo{year}{2005}),
  \urlprefix\url{https://link.aps.org/doi/10.1103/RevModPhys.77.137}.

\bibitem[{\citenamefont{Sompolinsky et~al.}(1988)\citenamefont{Sompolinsky,
  Crisanti, and Sommers}}]{Sompolinsky1988}
\bibinfo{author}{\bibfnamefont{H.}~\bibnamefont{Sompolinsky}},
  \bibinfo{author}{\bibfnamefont{A.}~\bibnamefont{Crisanti}}, \bibnamefont{and}
  \bibinfo{author}{\bibfnamefont{H.~J.} \bibnamefont{Sommers}},
  \bibinfo{journal}{Phys. Rev. Lett.} \textbf{\bibinfo{volume}{61}},
  \bibinfo{pages}{259} (\bibinfo{year}{1988}),
  \urlprefix\url{https://link.aps.org/doi/10.1103/PhysRevLett.61.259}.

\end{thebibliography}


\appendix

\section{Convergence of the population dynamics algorithm}
\label{append}

\begin{figure}[H]
  \begin{center}
    \includegraphics[scale=0.35]{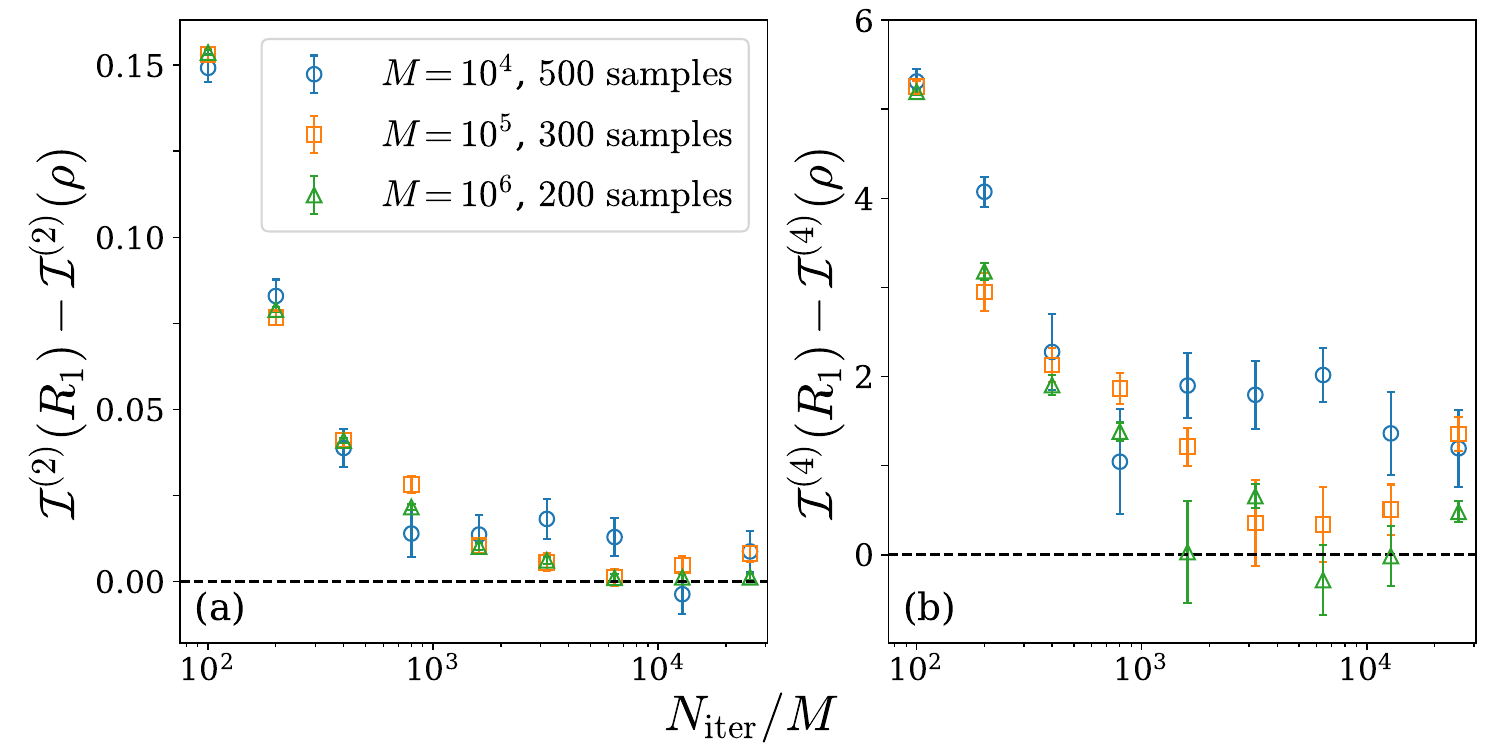} 
    \caption{Dimensionless second ($\mathcal{I}^{(2)}(\rho)$) and fourth ($\mathcal{I}^{(4)}(\rho)$) moments of the infection probabilities.
   The panels show the difference between the analytic results, $\mathcal{I}^{(2)}(R_1)$ and $\mathcal{I}^{(4)}(R_1)$ (see Eqs. (\ref{r1}) and (\ref{r2})), and 
      the numerical results for $\mathcal{I}^{(2)}(\rho)$ and $\mathcal{I}^{(4)}(\rho)$, obtained from the solutions of Eq. (\ref{hhjj2})
      using the population dynamics algorithm with population size $M$ and total number $N_{\rm iter}$ of iterations (vertical
      bars indicate the standard deviation of the mean values computed over different samples). These results are for
      parameters near the critical point in figure \ref{IPR}: Poisson indegrees, $\Gamma$-distribution
      of infection rates with mean $\lambda = 1/2$ and standard deviation $\sigma=0.3$, and $c-\lambda^{-1}=10^{-4}$.
}
\label{IPRappend}
\end{center}
\end{figure}

In this appendix, we present numerical results for the influence of both the population size $M$ and the number of iterations $N_{\rm iter}$ in the population dynamics
algorithm  near the transition point $\sigma_4 \simeq 0.36$ of figure \ref{IPR}. Figure \ref{IPRappend} shows 
the difference between the analytic results for $N \rightarrow \infty$, Eqs. (\ref{r1}) and (\ref{r2}), and the numerical results
for $\mathcal{I}^{(2)}(\rho)$ and $\mathcal{I}^{(4)}(\rho)$ obtained from the solutions of Eq. (\ref{hhjj2}).
For relatively small
number of iterations, there is a significant discrepancy between analytic and numerical results, which is not evident in figure \ref{IPR}
due to the logarithmic scale of the vertical axis. However, the data in figure \ref{IPRappend} indicate that, despite the fluctuations, the
numerical results consistently converge to the theoretical predictions as we increase both $M$ and $N_{\rm iter}$, giving further support
to the analytic curves in figure \ref{IPR}.
This critical slowing down of the population dynamics method close to $\sigma_4$ makes accurate calculations
of higher moments of $\mathcal{P}(\rho)$ computationally very expensive.

\end{document}